\documentclass{amsart}
\usepackage{amsmath, amssymb, amsthm, amscd, graphicx}
\usepackage{epstopdf}

\theoremstyle{plain}
\newtheorem{prop}{Proposition}[section]
\newtheorem{coro}[prop]{Corollary}

\newtheorem{lemm}[prop]{Lemma}

\theoremstyle{definition}
\newtheorem{defi}[prop]{Definition}
\newtheorem{nota}[prop]{Notation}
\newtheorem{exam}[prop]{Example}
\newtheorem{rema}[prop]{Remark}

\numberwithin{equation}{section}

\def\Reff#1; #2; #3; #4; #5; #6; #7\par{%
\bibitem{#1} #2, {\it #3}, #4 {\bf #5} (#6) #7}

\def\Ref#1; #2; #3; #4\par{%
\bibitem{#1} #2, {\it #3}, #4}

\let\a=\alpha
\renewcommand{\aa}[1]{a_{#1}}
\renewcommand{\AA}[1]{A_{#1}}
\newcommand{\AAm}[1]{A_{#1}^{\scriptscriptstyle-}}
\newcommand{\AAp}[1]{A_{#1}^{\scriptscriptstyle+}}
\newcommand{\act}{\cdot}
\newcommand{\ALD}{\mathtt{ALD}}
\newcommand{\ALDi}{\mathtt{ALD}_1}
\newcommand{\ALDii}{\mathtt{ALD}_2}
\newcommand{\Ass}{\mathtt{A}}

\let\b=\beta

\newcommand{\blue}[1]{\widehat{#1}}
\newcommand{\Bp}{B_{\scriptscriptstyle\bullet}}

\newcommand{\cf}{{\it cf.}\ }

\newcommand{\cl}{\mathtt{cl}}

\newcommand{\comp}{\mathbin{\scriptstyle\bullet}}

\let\d=\delta

\newcommand{\ea}{{\scriptstyle\varnothing}} % empty address

\newcommand{\ee}{e} % orientation 

\newcommand{\EQ}[1]{\mathrel{\leftrightarrow_{{#1}}}}
\newcommand{\etc}{{\it etc.}}
\newcommand{\ev}{\mathtt{eval}}
\newcommand{\eev}{\mathit{eval}}

\newcommand{\ff}{f} % a function

\let\g=\gamma
\let\ge=\geqslant
\newcommand{\GA}{\Geo{\mathtt{A}}}
\newcommand{\GALD}{\Geo\ALD}
\newcommand{\GALDp}{\Geo\ALD^{\scriptscriptstyle+}}
\renewcommand{\gg}{g} % a function
\newcommand{\Geo}[1]{\mathtt{Geom}_{#1}}
\newcommand{\GGeo}[1]{\mathcal{G}eom_{#1}}
\newcommand{\GGA}{\GGeo{\mathtt{A}}}
\newcommand{\GGAp}{\GGeo{\mathtt{A}}^{\scriptscriptstyle+}}
\newcommand{\GGALD}{\GGeo{\mathtt{ALD}}}
\newcommand{\GGALDp}{\GGeo{\mathtt{ALD}}^{\scriptscriptstyle+}}

\newcommand{\GGLL}{\GGeo{\LLL}}
\newcommand{\GGLLp}{\GGeo{\LLL}^{\scriptscriptstyle+}}

\newcommand{\GLD}{\Geo{\LD}}
\newcommand{\GLDp}{\Geo{\LD}^{\scriptscriptstyle+}}
\newcommand{\GLL}{\Geo{\LLL}}
\newcommand{\GLLp}{\Geo{\LLL}^{\scriptscriptstyle+}}

\newcommand{\id}{id}
\newcommand{\ie}{{\it i.e.}}

\newcommand{\IInt}{\mathcal{C}c}
\renewcommand{\Im}{\mathrm{Im}}
\newcommand{\Int}{\mathtt{Cc}}
\newcommand{\inv}{^{-1}}

\newcommand{\LD}{\mathtt{LD}}

\newcommand{\LL}{L}% algebraic law
\newcommand{\LLbis}{L'}% another algebraic law
\newcommand{\LLL}{\mathcal{L}}% family of algebraic laws
\newcommand{\OO}[1]{L_{#1}}%Operateur
\newcommand{\OOp}[1]{L_{#1}^{\scriptscriptstyle+}}%Operateur
\newcommand{\OObisp}[1]{{L'}_{#1}^{\scriptscriptstyle+}}%Operateur
\newcommand{\OOO}[2]{{#1}_{#2}}%Operateur
\newcommand{\OOOm}[2]{{#1}_{#2}^{\scriptscriptstyle-}}%Operateur
\newcommand{\OOOp}[2]{{#1}_{#2}^{\scriptscriptstyle+}}%Operateur
\newcommand{\op}{*}
\newcommand{\Op}{\mathbin{\scriptstyle\square}}
\newcommand{\OP}{\circ}

\newcommand{\rel}{\diamond}
\newcommand{\resp}{{\it resp.{~}}}

\newcommand{\RLL}{R_{\LLL}}
\newcommand{\RLLp}{R_{\LLL}^{\scriptscriptstyle+}}

\newcommand{\sh}[1]{\mathtt{sh}_{#1}}
\newcommand{\ssh}[1]{\mathit{sh}_{#1}}
\newcommand{\Sign}{\mathcal{F}} % signature
\renewcommand{\ss}[1]{\sigma_{#1}}
\renewcommand{\SS}[1]{\Sigma_{#1}}
\newcommand{\SSm}[1]{\Sigma_{#1}^{\scriptscriptstyle-}}
\newcommand{\SSp}[1]{\Sigma_{#1}^{\scriptscriptstyle+}}
\newcommand{\sub}[2]{\mathtt{sub}(#1,#2)} % subterm
\newcommand{\subst}{\sigma} % substitution
\newcommand{\substt}{\tau} % substitution

\newcommand{\Term}[1]{\mathtt{Term}_{#1}} % terms
\newcommand{\TTerm}[2]{\mathtt{Term}_{#1,#2}} % terms
\newcommand{\tm}{l}
\newcommand{\toLDp}{\to_{\LD}^{\scriptscriptstyle+}}
\newcommand{\toLLp}{\to_{\LLL}^{\scriptscriptstyle+}}

\newcommand{\Tox}{\Term{\op,\OP,\xx}}
\newcommand{\tp}{r}
\renewcommand{\tt}{t} % a term
\newcommand{\tti}{t_1} % a subterm
\newcommand{\ttii}{t_2} % a subterm
\newcommand{\ttiii}{t_3} % a subterm
\newcommand{\ttz}{t_0} % a subterm
\newcommand{\ttt}{t'} % another term
\newcommand{\ttti}{t'_1} % a subterm
\newcommand{\tttii}{t'_2} % a subterm
\newcommand{\tttt}{t''} % another term
\newcommand{\ttttt}{t'''} % another term
\newcommand{\tttz}{t'_0} % another subterm
\newcommand{\TS}{\Term\Sign} % terms
\newcommand{\TSV}{\TTerm\Sign\Var} % terms
\newcommand{\TSx}{\TTerm\Sign\xx}

\newcommand{\Var}{\mathcal{X}} % variables

\newcommand{\ww}{w}
\newcommand{\WW}[1]{\mathcal{W}_{#1}}
\newcommand{\WALD}{\mathcal{W}_{\ALD}}
\newcommand{\WLL}{\mathcal{W}_{\LLL}}

\newcommand{\xx}{x}
\newcommand{\XX}[1]{L_{#1}}
\newcommand{\XXp}[1]{L_{#1}^{\scriptscriptstyle+}}

\newcommand{\yy}{y}
\newcommand{\YY}[1]{L'_{#1}}
\newcommand{\YYp}[1]{{L'}_{#1}^{\scriptscriptstyle+}}

\newcommand{\zz}{z}

\begin{document}

\author{Patrick DEHORNOY}
\address{Laboratoire de Math\'ematiques Nicolas
Oresme UMR 6139\\ Universit\'e de Caen,
14032~Caen, France}
\email{dehornoy@math.unicaen.fr}
\urladdr{//www.math.unicaen.fr/\!\hbox{$\sim$}dehornoy}

\title{Using groups for investigating rewrite systems}



\begin{abstract}
We describe several technical tools that prove to be efficient for investigating the rewrite systems associated with a family of algebraic laws, and might be useful for more general rewrite systems. These tools consist in introducing a monoid of partial operators, listing the monoid relations expressing the possible local confluence of the rewrite system, then introducing the group presented by these relations, and finally replacing the initial rewrite system with a internal process entirely sitting in the latter group. When the approach can be completed, one typically obtains a practical method for constructing algebras satisfying prescribed laws and for solving the associated word problem.
\end{abstract}

\maketitle

\section*{Introduction}

Let~$\LLL$ be a family of algebraic laws, typically associativity, commutativity, distributivity, \etc, involving a certain signature~$\Sign$, and let~$\Var$ be an infinite set of variables. By associating with every law $\tm = \tp$ of~$\LLL$ the two rules $\tm \to \tp$ and $\tp \to \tm$, we obtain a rewrite system~$R_{\LLL}$ on the family of all $\Sign$-terms constructed on~$\Var$. The aim of this paper is to present a general method for investigating the rewrite systems of the form~$R_{\LLL}$ by introducing an associated monoid or group. This approach proved to be useful for various systems~$R_{\LLL}$, and it might be relevant for more general rewrite systems, typically those arising in a context of algebra~\cite{Coh, DeJ}.

The approach comprises three steps.
The first one consists in associating with the rewrite system~$R_{\LLL}$ a certain inverse monoid ~$\GGeo{\LLL}$ of partial operators by taking into account where and in which direction the rules are applied; this monoid is called the {\it geometry monoid} for~$\LLL$ as it captures several specific geometrical phenomena specific to~$\LLL$.

The second step consists in replacing the inverse monoid~$\GGeo{\LLL}$ with a group $\Geo{\LLL}$ that resembles it: when the laws of~$\LLL$ are simple enough, typically when no variable is repeated more than once in each side of the laws, the group~$\Geo{\LLL}$ can be defined to simply be the universal group of~$\GGeo{\LLL}$; in more complicated cases, a convenient group can be obtained by investigating the local confluence relations holding in~$\GGeo{\LLL}$ and defining~$\Geo{\LLL}$ to be the group presented by these relations. 

The third---and technically main---step is an internalization process that replaces the external action of~$\GGeo{\LLL}$ on terms by an internal multiplication in~$\Geo{\LLL}$. 

When all three steps can be completed, various questions about the laws~$\LLL$ can be successfully addressed, typically constructing algebras that satisfy these laws, or solving the word problem of~$\LLL$, \ie, designing an algorithm that decides whether two terms are equivalent under the congruence generated by~$R_{\LLL}$.

In this paper, we shall present the approach in a general setting and mention some of the existing examples. However, in order to make this paper more than a survey, we shall put the emphasis on a new example for which the method had not been considered before, namely the case of  the augmented left self-distributivity laws, or $\ALD$-laws, defined to be the following three laws
\begin{equation}
\label{E:ALD}
\tag{$\ALD$}
\begin{cases}
\ x \op (y \op z) = (x \op y) \op (x \op z),\\
\ x \op (y \OP z) = (x \op y) \OP (x \op z),\\
\ x \op (y \op z) = (x \OP y) \op z.
\end{cases}
\end{equation}
These laws and the algebras that satisfy them, called {\it $\ALD$-algebras} in this paper, appeared recently in several frameworks \cite{Dhb, Dhe, Dhj}, and they had never been addressed from the point of view of the geometry monoid. We shall see that a number of technical questions remain open in this seemingly difficult case. Nevertheless, the method is sufficient to naturally lead to the construction of a (highly non-trivial) example of such an $\ALD$-algebra. Technically, the main step is the construction of what is called a blueprint for the $\ALD$-laws.

The leading principle underlying the approach is to use the geometry monoid to {\it guess} some formulas, and then to check the latter by a direct verification. Typically, concentrating on the possible confluence relations that hold in the geometry monoid~$\GGLL$ and using these relations for introducing a group~$\GLL$ relies on a heuristic approach rather than on any uniform proved argument. Now the point is that this approach, which might seem loose, actually works in some definitely non-trivial cases: thus what legitimates the scheme is not some general a priori argument, but rather its a posteriori success.

The paper is organized as follows. The geometry monoid~$\GGLL$ is introduced in Section~\ref{S:Monoid}. The process for going from the monoid~$\GGLL$ to a group~$\GLL$ is described in Section~\ref{S:Group}. The principle of internalizing terms in the geometry monoid/group is explained in Section~\ref{S:Blue}. Next, we show in Section~\ref{S:Using} how the general study developed in Sections~\ref{S:Group} and~\ref{S:Blue} can be used, in good cases, to investigate a family of algebraic laws, with a special emphasis on the example of~$\ALD$. Finally, in Section~\ref{S:Conf}, we briefly address the question of recognizing whether~$\GLL$ is a group of fractions, which amounts to investigating whether a locally confluent system is actually confluent: we insist on some methods for doing that even in a non-noetherian framework, \ie, when the standard methods fail.

\section{The geometry monoid~$\GGLL$}
\label{S:Monoid}

The first step in our study consists in analysing a rewrite system by means of a monoid of partial operators. The general idea is as follows. If $R$ is a rewrite system on some set~$T$, then, in the most general situation, several rules may be applied to a given element~$\tt$ of~$T$, and, on the other hand, not every rule need to apply to~$\tt$, so the action of~$R$ on~$T$ cannot be described by functional operators in a natural way. However, by fixing the value of enough parameters, we can always discard the lack of uniqueness and describe~$R$ in terms of partial functional operators. 

Here we shall apply this scheme in the case when $T$ is a set of terms and $R$ consists of rules that can be applied to various subterms, more precisely when $R$ is the rewrite system associated with a family of algebraic laws. In this case, we shall obtain uniqueness, hence functionality, whenever the rule and the position are fixed.

\subsection{Applying an algebraic law}

For $\Sign$ a signature consisting of operation symbols exclusively, and $\Var$ a nonempty set of variables, we denote by $\TSV$ the family of all terms built using operation symbols from~$\Sign$ and variables from~$\Var$. Practically, we shall assume that some infinite list of variables has been fixed, and drop the reference to~$\Var$, thus writing~$\Term\Sign{}$ for~$\TSV$. An algebraic law is a pair of terms $\{\tm, \tp\}$---usually denoted as $\tm = \tp$. We shall speak of an {\it ordered} algebraic law to insist that the pair~$(\tm, \tp)$ is ordered, \ie, a distinguished orientation is chosen. For instance, there are two oriented versions for the standard associativity law $\xx \cdot (\yy \cdot \zz) = (\xx \cdot \yy) \cdot \zz$, namely the pair $(\xx \cdot (\yy \cdot \zz), (\xx \cdot \yy) \cdot \zz)$, and the symmetric pair. Note that, as the commutativity law is $\xx \cdot \yy = \yy \cdot \xx$ is involutive, there is, up to a substitution of variables, only one oriented version.

\begin{figure} [htb]
\begin{picture}(70,42)
\put(17,39){$\tt$}
\put(58,39){$\ttt$}
\put(0,7){\includegraphics{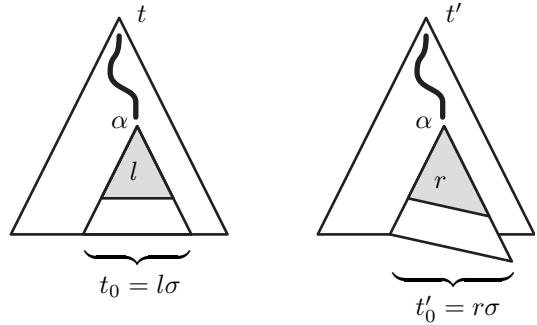}}
\put(16,18){$\tm$}
\put(13.5,25){$\a$}
\put(54,25){$\a$}
\put(10,9){$\underbrace{\hbox to 14mm{\hfill}}$}
\put(12,3){$\ttz=\tm\subst$}
\put(56.5,17){$\tp$}
\put(51,6){$\underbrace{\hbox to 16mm{\hfill}}$}
\put(54,0){$\tttz=\tp\subst$}
\end{picture}
\caption{\smaller Applying the law $\tm=\tp$ to a term~$\tt$, here viewed as a rooted tree: one replaces some subterm~$\ttz$ of~$\tt$ that is a substitute of~$\tm$ with the corresponding substitute of~$\tp$---or vice versa.}
\label{F:Apply}
\end{figure}

In this context, the fundamental operation of {\it applying} an algebraic law to a term corresponds to a rewrite system: applying $\tm=\tp$---\ie, $\{\tm, \tp\}$---to some term~$\tt$ means replacing some subterm~$\ttz$ of~$\tt$ which happens to be a substitute of~$\tm$ with the corresponding substitute~$\tttz$ of~$\tp$, or {\it vice versa}. This means that there exists a position~$\a$ and a substitution~$\subst$, \ie, a mapping from~$\Var$ into~$\TS$, such that the $\a$th subterm of~$\tt$ is $\tm\subst$ and $\ttt$ is obtained from~$\tt$ by replacing the $\a$th subterm with~$\tp\subst$---or {\it vice versa} exchanging the roles of~$\tm$ and~$\tp$ (Figure~\ref{F:Apply}). In other words, we apply one of the two rules $\tm \to \tp$, $\tp \to \tm$.

\begin{defi}
For each family of algebraic laws~$\LLL$ involving the signature~$\Sign$, we denote by~$R_{\LLL}$ the rewrite system on~$\Term\Sign{}$ consisting of all rules $\tm \to \tp$ and $\tp \to \tm$ for $\tm = \tp$ a law of~$\LLL$.
\end{defi}

Thus, by very definition, we have

\begin{prop}
\label{P:Equiv}
Let $\LLL$ be a family of algebraic laws involving the signature~$\Sign$. Then two terms~$\tt, \ttt$ in~$\TS$ are $\LLL$-equivalent if and only if one has $\tt \EQ\LLL \ttt$, where $\EQ\LLL$ is the equivalence relation generated by~$R_{\LLL}$.
\end{prop}

\subsection{Geometry monoid: the principle}

The rewrite systems~$R_{\LLL}$ are not functional: there may be many ways to apply one of its rules to a given term. However, these systems can be viewed as the union of a family of partial functions, each of which corresponds to choosing a law, an orientation, and a position.

For the sequel, it is convenient to fix an addressing system for the subterms of a term, \ie, for the positions in a term. As is usual, we shall see terms as rooted trees (\cf~Figure~\ref{F:Apply}), where the inner nodes are labeled using operation symbols, and the leaves are labeled using variables. Then a subterm of a term~$\tt$ is naturally specified by the node where its root lies, which is itself determined by the path that connects the root of the tree to that node. If all operation symbols are binary---which will be the case in the examples considered below---we can for instance use finite sequences of~$0$'s and~$1$'s to describe such a path, using $0$ for ``forking to the left'' and $1$ for ``forking to the right''. We use $\ea$ for the empty address, \ie, the address of the root. If $\tt$ is a term, and $\a$ is an address, we denote by $\sub\tt\a$ the $\a$-th subterm of~$\tt$, \ie, the subterm of~$\tt$ whose position is specified by~$\a$. Note that $\sub\tt\a$ exists only for~$\a$ short enough: $\sub\tt\ea$ always exists and equals~$\tt$, but, for instance, $\sub\tt0$ and~$\sub\tt1$, which are the left and the right subterms of~$\tt$ respectively, exist only if $\tt$ is not a variable.

\begin{defi}\label{D:Oper}
$(i)$ Assume that $\LL$ is an oriented algebraic law. For each address~$\a$ and each orientation~$\ee$ (namely $+$ or $-$), we denote by $\OOO\LL\a^\ee$ the (partial) operator on terms corresponding to applying~$\LL$ at
position~$\a$ in direction~$\ee$. 

$(ii)$ For~$\LLL$ a family of oriented laws, we define the {\it geometry monoid of~$\LLL$}, denoted~$\GGLL$, to be the monoid generated, using composition, by all operators~$\OOO\LL\a^\ee$ when $\LL$ ranges over~$\LLL$, $\a$ ranges over all addresses, and $\ee$ ranges over~$\{+,-\}$.
\end{defi}

We shall always think of the operators~$\OOO\LL\a^\ee$ as acting on the right, thus writing $\tt \act \ff$ rather than $\ff(\tt)$ for the image of the term~$\tt$ under the operator~$\ff$. To be coherent, we use the reversed composition, denoted~$\comp$, in the geometry monoid. Thus, $\ff \comp \gg$ means ``$\ff$ then $\gg$''.

\begin{rema}
When a term~$\ttz$ is a substitute of the term~$\tm$, the substitution~$\subst$ such that $\ttz = \tm\subst$ is not unique, as the value of~$\xx\subst$ is uniquely determined only for those variables~$\xx$ that occur in~$\tm$. Hence, for $\LL = (\tm, \tp)$, the operator~$\OOOp\LL\a$ is functional only if the same variables occur in~$\tm$ and~$\tp$. The laws that satisfy this condition will be called {\it balanced}. Although this is not necessary, we shall always restrict to balanced laws in the sequel. Thus we discard laws like $\xx = \xx \op \yy$; note that the algebras obeying such laws are trivial.
\end{rema}

By construction, every element in a geometry monoid~$\GGLL$ is a finite product of operators~$\OOO\LL\a^e$ with~$\LL$ in~$\LLL$. It will be convenient in the sequel to fix the following notation:

\begin{nota}
For~$\LLL$ a family of (oriented, balanced) laws, we denote by~$\WLL$ the free monoid generated by all letters~$\OOO\LL\a^e$ with~$\LL \in \LLL$, $\a \in \{0,1\}^*$, and $e \in \{+,-\}$; we denote by $\eev$ the canonical evaluation morphism 
$$\eev : \WLL \to \GGLL.$$
\end{nota} 

Thus $\WLL$ consists of all abstract products of letters~$\OOO\LL\a^e$, while $\GGLL$ consists of actual operators on terms. We shall see soon that, in general (and as can be expected), the evaluation mapping is far from injective, \ie, the geometry monoid~$\GGLL$ is far from free.

\subsection{Geometry monoid: the example of~$\ALD$}
\label{S:ALD}

To illustrate our approach, we consider the family~$(\ALD)$ consisting of the following three algebraic laws:
\begin{gather}
\tag{$\LD$}\label{E:LD}
\xx \op (\yy \op \zz) = (\xx \op \zz) \op (\xx \op \zz),\\
\tag{$\ALDi$}\label{E:ALD1}
\xx \op (\yy \OP \zz) = (\xx \op \zz) \OP (\xx \op \zz),\\
\tag{$\ALDii$}\label{E:ALD2}
\xx \op (\yy \op \zz) = (\xx \OP \yy) \op \zz.
\end{gather}
The specific interest of this choice is that, contrary to~$(\LD)$ alone, the above mixed laws, collectively denoted~$(\ALD)$---``Augmented Left self-Distributivity''---in the sequel, have never been investigated from the viewpoint of the geometry monoid and, so, the results we shall obtain below are new.

Here the signature consists of two binary operation symbols~$\op, \OP$. The law~\eqref{E:LD} is the left self-distributivity law, which was extensively studied in~\cite{Dgd}. The additional laws~\eqref{E:ALD1} and~\eqref{E:ALD2} express that $\op$ is left distributive with respect to~$\OP$ and that $\OP$ behaves like a sort of composition relative to~$\op$. Many examples of $\LD$-algebras, \ie, of structures satisfying~\eqref{E:LD}, happen to be equipped with a second operation that satisfies the mixed laws~$(\ALDi)$ and~$(\ALDii)$. This is in particular the case for every group equipped with the $\LD$-operation $\xx \op \yy := \xx \yy\xx\inv$: in this case, defining the second operation to be the multiplication $\xx \OP \yy := \xx \yy$ yields an $\ALD$-algebra---and more, actually, namely an $\LD$-monoid in the sense of~\cite{Dgd}, Chapter~XI.

By definition, the geometry monoid~$\GGALD$ is generated by three families of operators, corresponding to the three laws. In the specific case, we observe that the actions of the operators $\OOOp\LD\a$ and $(\ALDi)\OOOp{}\a$ are parallel, as both consist in distributing the left subterm to the two halves of the right subterm. However, their domains are disjoint, as $\OOOp\LD\a$ applies only when the symbol at~$\a1$ is~$\op$, while $(\ALDi)\OOOp{}\a$ applies only when the symbol at~$\a1$ is~$\OP$. Hence, instead of considering two operators separately, we shall introduce their union, which is still functional, and denote it by~$\SSp\a$. In particular, we have
\begin{equation}
\SSp\ea : \tti \op (\ttii \Op \ttiii) \to (\tti \op \ttii)
\Op (\tti \op \ttiii), 
\end{equation}
\ie, $\SSp\ea$, which will be also simply denoted $\SSp{}$, is the operator that maps every term of the form $\tti \op (\ttii \Op \ttiii)$ to the corresponding term $(\tti \op \ttii) \Op (\tti \op \ttiii)$, where $\Op$ stands for either~$\op$ or~$\OP$. Similarly, we have
\begin{equation}
\AAp\ea: \tti \op (\ttii \op \ttiii) \to (\tti \OP \ttii) \op \ttiii,
\end{equation}
\ie, $\AAp\ea$, also denoted~$\AAp{}$, maps every term of the form $\tti \op (\ttii \op \ttiii)$ to $(\tti \OP \ttii) \op \ttiii$. By definition, the geometry monoid~$\GGeo\ALD$ is the monoid generated by all partial operators~$\SSp\a$, $\SSm\a$, $\AAp\a$, and~$\AAm\a$ with~$\a$ ranging over~$\{0,1\}^*$.

\begin{figure} [htb]
\begin{picture}(105,58)
\put(0,0){\includegraphics{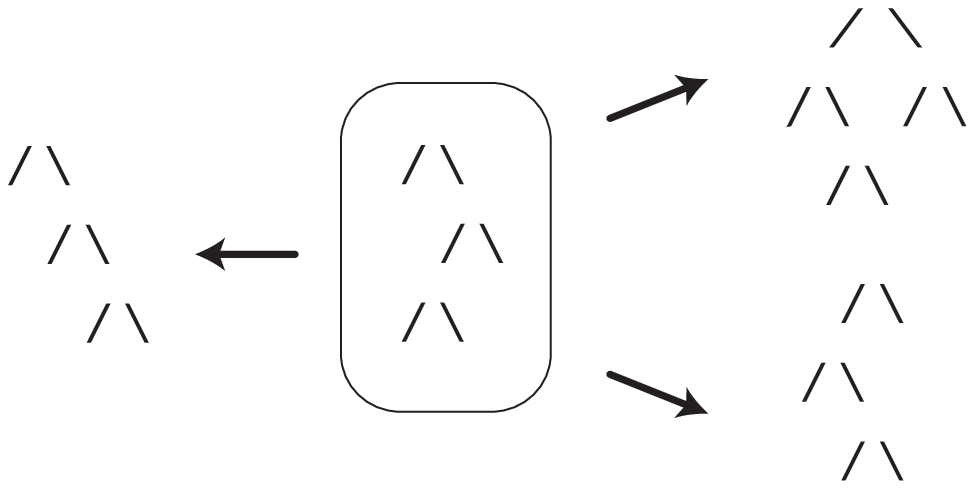}}
\put(5,40){$\op$}
\put(9,32){$\op$}
\put(1,32){$\xx_1$}
\put(13,24){$\op$}
\put(5,24){$\xx_2$}
\put(9,16){$\xx_3$}
\put(17,16){$\xx_4$}
\put(45,40){$\op$}
\put(49,32){$\op$}
\put(41,32){$\xx_1$}
\put(53,24){$\xx_4$}
\put(45,24){$\OP$}
\put(49,16){$\xx_3$}
\put(41,16){$\xx_2$}
\put(90,53){$\op$}
\put(84,45){$\op$}
\put(96,45){$\op$}
\put(80,37){$\xx_1$}
\put(88,37){$\OP$}
\put(92,37){$\xx_1$}
\put(100,37){$\xx_4$}
\put(84,29){$\xx_2$}
\put(92,29){$\xx_3$}
\put(90,25){$\op$}
\put(86,17){$\OP$}
\put(93,17){$\xx_4$}
\put(82,9){$\xx_1$}
\put(89.5,9){$\OP$}
\put(86,1){$\xx_2$}
\put(93,1){$\xx_3$}
\put(25,30){$\AA1^-$}
\put(67,16){$\AA\ea^+$}
\put(67,37){$\SS\ea^+$}
\end{picture}
\caption{\smaller Case of~$\ALD$: Two positive operators apply to the term $\xx_1 \op ((\xx_2 \OP \xx_3) \op \xx_4)$, namely~$\SSp\ea$ and~$\AAp\ea$, while one negative operator applies to it, namely~$\AAm1$, \ie, the copy of~$\AAm\ea$ acting at position~$1$.}
\label{F:Exam}
\end{figure}

As displayed in Figure~\ref{F:Exam}, a given term may belong to the domain and the range of several operators~$\SS\a^{\pm}$ and~$\AA\a^{\pm}$.

\begin{rema}
\label{R:Position}
For a signature~$\Sign$ comprising more than one operation symbol, as is the case with~$\ALD$, only taking into account the address where a law is applied does not exhaust all information. Indeed, when an operator~$\OO\a^e$ is applied to a term~$\tt$, the complete list of the operation symbols that occur in~$\tt$ above~$\a$ may be important, typically in terms of the relations possibly connecting various operators.
In order to include such data, we can use positions instead of addresses to index the operators of the geometry monoid, a {\it position} being defined to be a finite sequence of even length consisting of alternating operation symbols and forking directions: for instance, $(\op, 1)$ and $(\op, 1, \OP, 0)$ are positions that refine the addresses~$1$ and~$10$ respectively. In this framework, the operator~$\AA1^-$ mentioned in Figure~\ref{F:Exam} would become $\AA{(\op,1)}^-$, \ie, $\AA1^-$ applied to a term with~$\op$ at the root. Observe that an operator~$\OO\a^e$ as defined in Definition~\ref{D:Oper} is, as a set of pairs of terms, just the disjoint union of all operators~$\OO{p}^e$ with~$p$ a position that projects on the address~$\a$ when the operation symbols are forgotten. Although these refinements may be necessary in some cases, they are not in the examples considered in this paper, in particular~$\ALD$, and there will be no need to split the operators~$\OO\a^e$ into more elementary components.
\end{rema}

\subsection{Connection between $\GGLL$ and $\LLL$-equivalence}

We saw in Proposition~\ref{P:Equiv} that $\LLL$-equivalence is directly connected with the equivalence relation~$\EQ\LLL$ associated with the rewrite system~$R_{\LLL}$. As a consequence, it is also connected with the geometry monoid~$\GGLL$: 

\begin{prop}
\label{P:Basic}
Let $\LLL$ be any family of algebraic laws involving the signature~$\Sign$. Then two terms~$\tt, \ttt$ in~$\TS$ are $\LLL$-equivalent if and only if some element of the geometry monoid~$\GGLL$ maps~$\tt$ to~$\ttt$.
\end{prop}

\begin{proof}
Let us write $\tt \equiv \ttt$ if some operator of~$\GGLL$ maps~$\tt$ to~$\ttt$. First $\tt \equiv\ttt$ implies $\tt \EQ\LLL \ttt$. Indeed, by construction, $\ttt = \tt \act \OOO\LL\a^\ee$ implies $\tt \EQ\LLL \ttt$. As $\EQ\LLL$ is transitive, the same is true when several operators~$\OOO\LL\a^\ee$ are composed.

Conversely, $\EQ\LLL$ is the congruence generated by the instances of the laws of~$\LLL$, so, in order to prove that $\tt \EQ\LLL \ttt$ implies $\tt \equiv \ttt$, it is enough to prove that $\equiv$ is a congruence on terms, and that it contains all instances of the laws of~$\LLL$. Now, $\equiv$ is an equivalence relation because $\GGLL$ is closed under composition and contains the identity mapping, and it is a congruence, \ie, it is compatible with all operations of~$\Sign$. Indeed, assume that $\OOOp\LL\a$ maps~$\tt$ to~$\ttt$. Let $\ttz$ be any term and $\Op$ be any operation in~$\Sign$. Then---assuming that all operations of~$\Sign$ are binary and $0,1$~addresses are used---the operator $\OOOp\LL{0\a}$ maps~$\ttz\Op\tt$ to~$\ttz\Op\ttt$, while $\OOOp\LL{1\a}$ maps~$\tt\Op\ttz$ to~$\ttt\Op\ttz$. Finally, assume that $(\tt, \ttt)$ is an instance of some law~$\LL$ of~$\LLL$. This means that there exists a substitution~$\subst$ such that, assuming that $\LL$ is $\tm = \tp$, one has $\tt = \tm\subst$ and $\ttt = \tp\subst$. Now, by definition, $\OOOp\LL\ea$ maps~$\tt$ to~$\ttt$, so $\tt \equiv \ttt$ holds.
\end{proof}

Our aim in this paper is not to develop a general theory of the geometry monoid. However, we mention two results, namely one about the structure of~$\GGLL$ and one about its dependence on~$\LLL$.

First, we observe that, by construction, each operator in the geometry monoid~$\GGLL$ is close to admitting an inverse. Indeed, the operator~$\OOOm\LL\a$ is, as a set of pairs, the inverse of the operator~$\OOOp\LL\a$, \ie,  $\OOOp\LL\a$ maps~$\tt$ to~$\ttt$ if and only if $\OOOm\LL\a$ maps~$\ttt$ to~$\tt$. This is not enough to make~$\GGLL$ into a group, as the composition $\OOOp\LL\a \comp \OOOm\LL\a$ is the identity operator of its domain only, and {\it not} the identity operator of~$\TS$ in general, but we have:

\begin{prop}
Assume that $\LLL$ is a family of balanced algebraic laws. Then $\GGLL$ is an inverse monoid.
\end{prop}

\begin{proof}
Assume that $\ff$ is a nonempty operator in~$\GLL$. Then, by construction, there exists a word~$\ww$ in~$\WLL$ such that $\ff$ equals $\eev(\ww)$. Let $\ww\inv$ be the formal inverse of~$\ww$, \ie, the word obtained from~$\ww$ by exchanging~$\OOOp\LL\a$ and~$\OOOm\LL\a$ everywhere in~$\ww$ and reversing the order of the letters, and let $\gg := \eev(\ww\inv)$. An immediate induction shows that a pair of terms~$(\tt,\ttt)$ belongs to~$\ff$ if and only if the pair~$(\ttt,\tt)$ belongs to~$\gg$, and we deduce
$$\ff \comp \gg \comp \ff  = \ff
\mbox{\quad and \quad}
\gg \comp \ff \comp \gg = \gg,$$
so every element in~$\GLL$ possesses an inverse as required for an inverse monoid---this applies in particular to $\ff=\gg=\emptyset$.
\end{proof}

As for the second result, the {\it equational variety} associated with a family~$\LLL$ of algebraic laws involving the signature~$\Sign$ is, by definition, the collection of all $\Sign$-structures that satisfy the laws of~$\LLL$. Different families of laws may give rise to the same variety: for instance, when appended to the commutativity law $xy = yx$, the associativity law $x(yz) = (xy)z$ and the law $x(yz) = z(yx)$ define the same variety. 

\begin{prop}
\cite{Dew}
Up to isomorphism, the monoid~$\GGLL$ only depends on the equational variety defined by~$\LLL$: if $\LLL$ and $\LLL'$ define the same variety, then the monoids $\GGLL$ and $\GGeo{\LLL'}$ are isomorphic.
\end{prop}

\begin{proof} [Sketch of proof]
It suffices to show that, when we add to a family~$\LLL$ a new law~$\LL$ that is a consequence of the laws of~$\LLL$, then the geometry monoid is not changed. Assume $\LL = (\tm, \tp)$. By Proposition~\ref{P:Basic}, the hypothesis that $\LL$ is a consequence of~$\LLL$ implies that some operator~$\ff$ in~$\GGLL$ maps~$\tm$ to~$\tp$. The geometry monoid~$\GGeo{\LLL \cup \{\LL\}}$ is obtained from~$\GGLL$ by adding a new generator that is a product of the canonical generators of~$\GGLL$, and, therefore, it is isomorphic to~$\GGLL$.
\end{proof}

\section{Replacing the geometry monoid with a group}
 \label{S:Group}

The first specific step that proves to be often useful in investigating the geometry monoids consists in replacing the monoid~$\GGLL$ with a {\it group}~$\GLL$ that resembles it. The benefit of the replacement is the possibility of freely computing with inverses and avoiding the problems possibly connected with the empty operator. However, there is no universal recipe for going from~$\GGLL$ to a (non-trivial) group, and we shall complete the approach in specific cases only, by guessing confluence relations in~$\GGLL$ and introducing the abstract group defined by these relations.

\subsection{The linear case}
\label{SS:Linear}

For each inverse monoid~$M$, there exists a biggest quotient of~$M$ that is a group, namely the {\it universal group~$U(M)$} of~$M$ obtained by collapsing all idempotents to~$1$~\cite{Pat}. When the geometry monoid~$\GGLL$ does not contain the empty operator, the idempotent elements are those operators that are identity on their domain, and the associated group keeps enough information to be non-trivial. We shall first briefly mention one case when this favourable situation occurs, namely the case of {\it linear} laws.

\begin{defi}
A term~$\tt$ is said to be {\it injective} if no variable occurs twice in~$\tt$. An algebraic law $\tm = \tp$ is said to be {\it linear} if it is balanced and, in addition, the terms~$\tm$ and~$\tp$ are injective.
\end{defi}

For instance, associativity and commutativity are linear laws, while self-distributi\-vity is not, as the variable~$\xx$ is repeated twice in the right hand term of $\xx(\yy\zz) = (\xx\yy)(\xx\zz)$.

In order to describe the case of linear laws, we start with a general result about the elements of a geometry monoid.

\begin{defi}
Assume that $\ff$ is an operator on~$\TS$. We say that a pair of terms~$(\tm, \tp)$ is a {\it seed} for~$\ff$ if $\ff$, as a set of pairs, consists of all instances of~$(\tm, \tp)$, \ie, consists of all pairs~$(\tm \subst, \tp \subst)$ with~$\subst$ a $\TS$-valued substitution.
\end{defi}

If $\LL$ is the law $\tm = \tp$, then $\OOOp\LL\ea$ consists of all instances of~$(\tm, \tp)$, so $(\tm, \tp)$ is a seed for~$\OOOp\LL\ea$. More generally we have:

\begin{lemm} \cite{Dgd}
\label{L:Seed}
Assume that $\LLL$ is a family of balanced algebraic laws involving a single binary operation symbol. Then each nonempty operator~$\ff$ in~$\GGLL$ admits a seed.
\end{lemm}

\begin{proof} [Sketch of proof]
First, every operator~$\OOOp\LL\a$ admits a seed, because, if $(\tm,\tp)$ is a seed for~$\OOOp\LL\ea$ and $\xx$ is a variable not occurring in~$\tm$ and~$\tp$, then, assuming that $\op$ is the operation symbol, $(\xx \op\tm,\xx\op\tp)$ is a seed for~$\OOOp\LL1$, $(\tm\op\xx,\tp\op\xx)$ is a seed for~$\OOOp\LL0$, and an easy induction gives the result for every~$\OOOp\LL\a$.

Then, by construction, every operator in~$\GGLL$ is a finite composition of operators~$\OOO\LL\a^\ee$ with~$\LL$ in~$\LLL$, and, by definition, the latter admit seeds. Hence, the point is to show that the composition of two operators admitting a seed still admits a seed provided it is nonempty. So, assume that $\ff$ and~$\gg$ consist of all instances of~$(\tm_1, \tp_1)$ and~$(\tm_2, \tp_2)$, respectively. Two cases may occur. Either there exists no term that is a substitute both of~$\tp_1$ and~$\tm_2$, and in this case $\ff \comp \gg$ is empty. Or there exists such a term. Then, as is well known \cite{Rob}, there exists a most general unifier (MGU) of the terms~$\tp_1$ and~$\tm_2$, \ie, there exist two substitutions~$\subst, \substt$ satisfying $\tp_1 \subst = \tm_2 \substt$ and such that every common substitute of~$\tp_1$ and~$\tm_2$ is a substitute of~$\tp_1 \subst$. In this case, it is easy to check that $(\tm_1 \subst, \tp_2 \substt)$ is a seed for~$\ff \comp \gg$.
\end{proof}

\begin{rema}
The restriction on the signature in Lemma~\ref{L:Seed} can be dropped at the expense of splitting the operators~$\OOOp\LL\a$ according to positions as explained in Remark~\ref{R:Position}, or, alternatively, of considering a more general notion of terms and instances in which substitution is possible not only for variables but also for some specific operation symbols considered as variables of a new type. Once again, there is no need to go into details here, as we shall not use such notions. 
\end{rema}

\begin{lemm}
Assume that $\LLL$ is a family of linear laws involving a single binary operation symbol. Then the seed of every operator in~$\GGLL$ is a pair of injective terms. Moreover, $\GGLL$ does not contain the empty operator.
\end{lemm}

\begin{proof}
The MGU of two injective terms always exists and it is an injective term: indeed, the only reason that may cause an unifier not to exist is a variable clash, and this cannot happen with injective terms. So the result about the seed follows from an induction. As for the empty operator, it cannot appear as the needed unifier always exists.
\end{proof}

\begin{defi}
\label{D:Sim}
Assume that $\ff, \gg$ are partial mappings on some set~$S$. We say that $\ff \sim \gg$ (\resp $\ff\approx\gg$) holds if there exists at least one
element~$\xx$ in~$S$ such that $\ff(\xx) = \gg(\xx)$ holds (\resp\ if $\ff(\xx) = \gg(\xx)$ holds for every~$\xx$ in the intersection of the domains of~$\ff$ and~$\gg$).
\end{defi}

When the geometry monoid~$\GGLL$ contains the empty operator, the relations~$\sim$ and~$\approx$ need not be transitive. This however cannot happen in the linear case:

\begin{prop} \cite{Dhb} \label{P:Group}
\label{P:Geom}
Assume that $\LLL$ is a family of linear laws. Then the relations~$\sim$ and~$\approx$ coincide, and they are congruences on the monoid~$\GGLL$. Furthermore, the quotient-monoid~$\GGLL/\!\approx$ is a group, and it is the universal group of~$\GGLL$.
\end{prop}

Under the previous hypotheses, we shall denote by~$\GLL$ the
group~$\GGLL/\!\approx$. So, in this case, we have a scheme of the form
\begin{equation}
\label{E:Quotient}
\begin{CD}
\WLL
@>{\mathrm{onto}}>>
\GGLL
@>{\mathrm{onto}}>>
\GLL.
\end{CD}
\end{equation}
Moreover, the partial action of~$\GGLL$ on terms induces a well-defined action of~$\GLL$ as, by definition, all operators in a $\approx$-class agree on the terms that lie in their domain. Furthermore, no information is lost when we replace~$\GGLL$ with~$\GLL$ in the precise sense that the counterpart of Proposition~\ref{P:Basic} is true: two terms~$\tt, \ttt$ are $\LLL$-equivalent if and only if we have $\ttt = \tt \act \gg$ for some~$\gg$ in~$\GLL$. So, in this case, one can replace the monoid~$\GGLL$ with the group~$\GLL$ in all further uses, and it is then natural to call the latter the {\it geometry group} of~$\LLL$.

\begin{exam}
Let $\Ass$ denote the associativity law $x(yz) = (xy)z$. Then the corresponding group~$\GA$ is a well known group, namely R.\,Thompson's group~$F$ \cite{Tho}. In the case of associativity together with commutativity, the corresponding group is R.\,Thompson's group~$V$---\cf~\cite{Dhb}. 
\end{exam}

\begin{rema}
Even in the smooth case of linear laws, the action of the group~$\GLL$ on terms is a partial action: for $\tt \act \gg$ to be defined, it is necessary that $\tt$ be large enough. However, we can obtain an everywhere defined action by considering infinite trees; equivalently, this amounts to defining an action on the boundary of the family of finite terms, which is a Cantor set in the case of terms involving one binary operation symbol and one variable.
\end{rema}

\subsection{Confluence relations: the principle}

Whenever the empty operator occurs in the geometry monoid, the previous approach badly fails:

\begin{lemm}
\label{L:Empty}
Assume that $\GGLL$ contains the empty operator. Then the universal group $U(\GGLL)$ is trivial, \ie, it reduces to $\{1\}$.
\end{lemm}

\begin{proof}
Assume that $\pi$ is a homomorphism of~$\GGLL$ to a group. Then, for each~$\ff$ in~$\GGLL$, we have $\emptyset \comp \ff = \emptyset$, hence $\pi(\ff) \pi(\emptyset) = \pi(\emptyset)$, whence $\pi(\ff) = 1$.
\end{proof}

\begin{exam}
\label{X:Empty}
The case when there is no unifier frequently occurs. For instance, in the case of the self-distributivity law~$(\LD)$, every term~$\tt$ belonging to the range of~$\SSp\ea$ satisfies $\sub\tt{00} = \sub\tt{10}$; it follows that every term~$\tt$ in the range of~$\SSp\ea \comp \SSp1$ satisfies $\sub\tt{00} = \sub\tt{100}$, and therefore $\sub\tt{00} \not= \sub\tt{10}$. Hence no term in the image of $\SSp\ea \comp \SSp1$ mays belong to the domain of~$\SSm\ea$: in other words, the composition of the operators $\SSp\ea$, $\SSp1$, and $\SSm\ea$ is empty, \ie, the operator $\SSp{} \comp \SSp1 \comp \SSm{}$, as a set of ordered pairs, is the empty set, and, in particular, its domain of definition is empty. In other words, we have in~$\GGeo\LD$---as well as in~$\GGeo\ALD$---the relation
$$\eev(\SSp\eaÊ\SSp1 \SSm\ea) = \emptyset.$$
This implies that {any} word~$\ww$ in~$\WW\LD$ containing $\SSp\ea \SSp1 \SSm\ea$ as a factor evaluates in~$\GGeo\LD$ into the empty operator. This is in particular the case for the symmetrized word $\SSp\ea  \SSp1 \SSm\ea \SSp\ea \SSm1 \SSm\ea$, although the latter appears as freely reducing to the empty word and one might therefore expect the associated operator to be close to the identity.
\end{exam}

The above situation almost always occurs when non-linear laws are involved. As the example of $(xy)(xz) = (xz)(xy)$ shows, it is not readily true that the presence of at least one non-linear law in~$\LLL$ forces the empty operator to belong to~$\GGLL$, but, for instance, it is sufficient for implementing the argument used in Example~\ref{X:Empty} that $\LLL$ contains a law $\tm = \tp$ such that, for some variable~$\xx$, the sets $\{\a_1, ..., \a_p\}$ and $\{\b_1, ..., \b_q\}$ where~$\xx$ respectively occurs in~$\tm$ and~$\tp$ are distinct and at least one of them is not a singleton.

In such cases, the universal group~$U(\GGLL)$ is of no use, and we have to look for another method. A misleading attempt would be to try to modify the construction of the geometry monoid so as to artificially discard the empty operator. We doubt that anything interesting can occur by doing so---see \cite{Dgd} for a more thorough discussion. Instead, we shall now develop a completely different method for associating with~$\GGLL$ a group that keeps the meaningful information of~$\GGLL$, namely finding a presentation of~$\GGLL$ and introducing the group that admits this presentation---whatever its connection with~$\GGLL$ is.

Actually, finding a presentation of~$\GGLL$ is in general out of reach, at least by a direct approach. So, once again, we use an indirect approach consisting in isolating {\it some} relations satisfied in~$\GGLL$ using a uniform scheme, but not trying to prove that these relations make a presentation: the latter will possibly come at the very end when further constructions have been performed.

So, at this point, the problem is to find relations connecting the various operators~$\OOO\LL\a^\ee$ of~$\GGLL$. In the sequel, we shall concentrate  on {\it positive} relations, \ie, on relations that involve the operators~$\OOp\a$ but not their inverses.

\begin{defi}
For~$\LLL$ a family of oriented algebraic laws, we denote by~$\RLLp$ the rewrite system comprising the rules $\tm \to \tp$ for $(\tm, \tp)$ in~$\LLL$; the {\it positive geometry monoid}~$\GGLLp$ of~$\LLL$ is defined to be the submonoid of~$\GGLL$ generated by all operators~$\OOOp\LL\a$ with~$\LL$ in~$\LLL$ and~$\a$ in~$\{0,1\}^*$.
\end{defi}

The connection between the rewrite system~$\RLLp$ and the positive geometry monoid~$\GGLLp$ is similar to the connection between the rewrite system~$\RLL$ and the geometry monoid~$\GGLL$. We shall be looking for relations in~$\GGLLp$, \ie, for relations connecting the operators~$\OOp\a$ in~$\GGLLp$. The principle will be to investigate the relations that possibly arise when two operators are applied to one and the same term~$\tt$, which amounts to inverstigating the local confluence of the rewrite system~$\RLLp$. This means that, for all laws~$\LL, \LLbis$ in~$\LLL$ and all addresses~$\a, \b$, we look for relations of the generic form
\begin{equation}
\label{E:Conf}
\OOp\a \comp ... = \OObisp\b \comp ...
\end{equation}
In other words, we look for common right multiples in~$\GGLLp$.

\begin{defi}
For $\ff$ a (partial) mapping on~$\TS$ and $\a$ an address, we denote by~$\ssh\a(\ff)$ the {\it $\a$-shift} of~$\ff$, defined to be the partial operator on~$\TS$ that consists in applying~$\ff$ to the $\a$th subterm of its argument: $\tt \act \ssh\a(\ff)$ is defined if and only if $\sub\tt\a$ exists and $\sub\tt\a \act \ff$ is defined, and, in this case, $\tt \act \ssh\a(\ff)$ is obtained from~$\tt$ bt replacing the $\a$th subterm by its image under~$\ff$.
\end{defi}

So, for instance, we have $\OOOp\LL\a = \ssh\a(\OOOp\LL\ea)$ for each law~$\LL$ and each address~$\a$, and, more generally, $\OOOp\LL{\a\b} = \ssh\a(\OOOp\LL\b)$ for all~$\a, \b$.

As for confluence relations in the monoid~$\GGLLp$, two general schemes will be involved. The first one relies on the following general principle: 

\begin{lemm}
Assume that $\a$ and $\b$ are incomparable addresses, \ie, there exists
$\g$ such that $\g0$ is a prefix of~$\a$ and $\g1$ is a prefix of~$\b$,
or vice versa. Then, for all partial operators~$\ff, \gg$, we have
\begin{equation}
\label{E:Comm0}
\ssh\a(\ff) \comp \ssh\b(\gg) = \ssh\b(\gg) \comp \ssh\a(\ff).
\end{equation}
\end{lemm}

\begin{figure} [htb]
\begin{picture}(40,40)
\put(0,0){\includegraphics{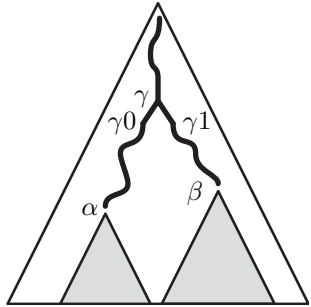}}
\put(10,12){$\a$}
\put(24,14){$\b$}
\put(17,27){$\g$}
\put(13.5,23.5){$\g0$}
\put(23.5,23.5){$\g1$}
\end{picture}
\caption{\smaller Incomparable addresses: The $\a$th and $\b$th subterms are disjoint, and therefore operators acting on them commute.}
\label{F:Ortho}
\end{figure}

\begin{proof}
(Figure~\ref{F:Ortho}) 
The operators~$\ssh\a(\ff)$ and~$\ssh\b(\gg)$ act on disjoint
subtrees, and therefore they commute. 
\end{proof}

In particular, we deduce:

\begin{prop}
Assume that $\a$ and $\b$ are incomparable addresses, \ie, there exists
$\g$ such that $\g0$ is a prefix of~$\a$ and $\g1$ is a prefix of~$\b$,
or conversely. Then we have
\begin{equation}
\label{E:Comm}
\OOp\a \comp \OObisp\b = \OObisp\b \comp \OOp\a.
\end{equation}
\end{prop}

The second general scheme appears in connection with what can be called
a {\it geometric inheritance} phenomenon.

\begin{lemm}
\label{L:Conf}
Assume that $\LL$ is the oriented law $(\tm, \tp)$ and that some variable~$\xx$ occurs at addresses $\b_1, ..., \b_p$ in~$\tm$ and at $\g_1, ..., \g_q$ in~$\tp$. Then, for each (partial) operator~$\ff$ acting on terms, we have
\begin{equation}
\label{E:Heir0}
\ssh{\b_1}(\ff) \comp ... \comp \ssh{\b_p}(\ff) \comp
\OOp\ea =
\OOp\ea \comp \ssh{\g_1}(\ff) \comp ... \comp
\ssh{\g_q}(\ff).
\end{equation}
More generally, for each address~$\a$, we have
\begin{equation}
\label{E:Heir1}
\ssh{\a\b_1}(\ff) \comp ... \comp \ssh{\a\b_p}(\ff) \comp
\OOp\a =
\OOp\a \comp \ssh{\a\g_1}(\ff) \comp ... \comp
\ssh{\a\g_q}(\ff).
\end{equation}
\end{lemm}

\begin{figure} [htb]
\begin{picture}(62,42)(0,-3)
\put(0,0){\includegraphics{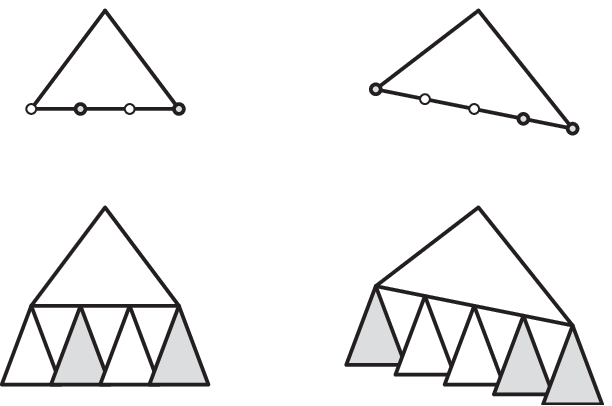}}
\put(0,0){\includegraphics{Heir.eps}}
\put(7,27){$\xx$}
\put(7,32){$\b_1$}
\put(17.5,32){$\b_p$}
\put(36.5,34){$\g_1$}
\put(57.5,30){$\g_q$}
\put(17,27){$\xx$}
\put(37,29){$\xx$}
\put(52,26){$\xx$}
\put(57,25){$\xx$}
\put(12,40){$\tm$}
\put(50,40){$\tp$}
\put(27,34){$\to$}
\put(27,15){$\to$}
\put(12,20){$\tm \subst$}
\put(50,20){$\tp \subst$}
\put(6,-0.5){$\xx \subst$}
\put(16,-0.5){$\xx \subst$}
\put(36,1.5){$\xx \subst$}
\put(51,-1.5){$\xx \subst$}
\put(56,-2.5){$\xx \subst$}
\put(7,12){$\b_1$}
\put(17.5,12){$\b_p$}
\put(36.5,14){$\g_1$}
\put(57.5,10){$\g_q$}
\end{picture}
\caption{\smaller Inheritance phenomenon: If the variable~$\xx$ occurs at~$\b_1, ..., \b_p$ in~$\tm$ and at $\g_1, ..., \g_q$ in~$\tp$ and nowhere else, then, for each term~$\tm\subst$, the subterm~$\xx\subst$ occurs at $\b_1, ..., \b_p$ in~$\tm\subst$, and t $\g_1, ..., \g_q$ in~$\tp\subst$, and applying~$\ff$ in each $\xx\subst$ before or after applying $\tm \to \tp$ leads to the same result.}
\label{F:Heir}
\end{figure}

\begin{proof}
(Figure~\ref{F:Heir})
Assume that $\OOp\ea$ maps~$\tt$ to~$\ttt$. This means that there exists a substitution~$\subst$ such that $\tt$ is~$\tm \subst$ and $\ttt$ is~$\tp \subst$. Let~$\subst_1$ be the substitution defined by $\yy\subst_1 = \yy \subst$ for $\yy \not= \xx$, and $\xx \subst_1 = \xx \subst \ff$. Let $\tt_1 := \tm \subst_1$ and $\ttt_1:= \tp \subst_1$. Then, by construction, $\OOp{}$ maps~$\tt_1$ to~$\ttt_1$. Now, $\tt_1$ is obtained from~$\tt$ by replacing the subterms at addresses $\b_1$, ..., $\b_p$ with their image under~$\ff$, so we have
$$\tt_1 = \tt \act (\ssh{\b_1}(\ff) \comp ... \comp \ssh{\b_p}(\ff)).$$
Similarly, $\ttt_1$ is obtained from~$\ttt$ by replacing the subterms at addresses $\g_1$, ..., $\g_q$ with their image under~$\ff$, so we have
$$\ttt_1 = \ttt \act (\ssh{\g_1}(\ff) \comp ... \comp \ssh{\g_q}(\ff)),$$
and \eqref{E:Heir0} follows.

Relation~\eqref{E:Heir1} is deduced by applying~$\ssh\a$ to both terms of~\eqref{E:Heir0}.
\end{proof}

By applying Lemma~\ref{L:Conf} to the case when the operator~$\ff$ has the form~$\OObisp\d$, we obtain:

\begin{prop}
Assume that $\LL$ is the oriented law $(\tm, \tp)$ and that some variable~$\xx$ occurs at addresses $\b_1, ...\b_p$ in~$\tm$ and at $\g_1, ..., \g_q$ in~$\tp$. Then, for all addresses~$\a, \d$ and each law~$\LLbis$, we have
\begin{equation}
\label{E:Heir}
\OOp\a \comp \OObisp{\a\g_1\d} \comp ... \comp
\OObisp{\a\g_q\d} =
\OObisp{\a\b_1\d} \comp ... \comp \OObisp{\a\b_p\d} \comp
\OOp\a.
\end{equation}
\end{prop}

In general, the previous two general schemes do not exhaust all possible confluence relations in~$\GGLL$. Typically, no relation is obtained for pairs of the form $(\a, \a\b)$ where $\b$ is so short that it falls inside the terms~$\tm$ or~$\tp$ associated with the considered law. In this case, no general scheme is known, and one has to look at the specific terms in order to find possible confluence relations.

\subsection{Confluence relations: the example of $\ALD$}
\label{SS:ConfALD}

We illustrate the previous scheme on the case of~$\ALD$. The first family of confluence relations comprises the trivial commutation relations, here all relations
\begin{equation}
\label{E:ALDComm}
\XX\a \comp \YY\b = \YY\b \comp \XX\a
\mbox{\quad for $\XX{}, \YY{} = \SS{}, \AA{}$ and $\a, \b$ incomparable}.
\end{equation}
The second family of confluence relations comprises the geometric inheritance relations. As for the laws~$\LD$ and~$\ALDi$, \ie, for the operator~$\SS{}$, three variables are involved in the rule 
\begin{equation}
\label{E:Sigma}
x \op (y \Op z) \to (x \op y) \Op (x \op z).
\end{equation}
The variable~$x$ occurs at address~$0$ on the LHS of~\eqref{E:Sigma}, while it occurs at~$00$ and~$10$ on the RHS. So Relation~\eqref{E:Heir} is here
\begin{equation}
\label{E:ALDHeir1}
\SSp\a \comp \XXp{\a0\d} = \XXp{\a00\d} \comp \XXp{\a10\d} \comp \SSp\a
\end{equation}
for $\XX{} = \SS{}, \AA{}$. Similarly, the variable~$y$ occurs at~$10$ on the LHS of~\eqref{E:Sigma}, and at~$01$ on the RHS, while the variable~$z$ occurs at~$11$ on both sides. Thus  Relations~\eqref{E:Heir} become
\begin{gather}
\label{E:ALDHeir2}
\SSp\a \comp \XXp{\a10\d} = \XXp{\a01\d} \comp \SSp\a, \\
\label{E:ALDHeir3}
\SSp\a \comp \XXp{\a11\d} = \XXp{\a11\d} \comp \SSp\a
\mbox{\quad for $\XX{} = \SS\ea, \AA\ea$}.
\end{gather}
The treatment is analogous for the law~$\ALDii$, \ie, for the operator~$\AA{}$. Three
variables occur in the rule
\begin{equation}
\label{E:A}
x \op (y \op z) \to (x \OP y) \op z.
\end{equation}
The variable~$x$ occurs at~$0$ on the LHS of~\eqref{E:A}, and at~$00$ on the RHS; $y$ occurs at~$10$ and at~$01$ respectively; finally, $z$ occurs at~$11$ and at~$1$. The corresponding Relations~\eqref{E:Heir} are
\begin{gather}
\label{E:ALDHeir4}
\AAp\a \comp \XXp{\a0\d} = \XXp{\a00\d} \comp \AAp\a, \\
\label{E:ALDHeir5}
\AAp\a \comp \XXp{\a10\d} = \XXp{\a01\d} \comp \AAp\a, \\
\label{E:ALDHeir6}
\AAp\a \comp \XXp{\a11\d} = \XXp{\a1\d} \comp \AAp\a,
\end{gather}
again for $\XX{} = \SS{}, \AA{}$. 

\begin{figure} [htb]
\begin{picture}(74,48)
\put(0,0){\includegraphics[scale=0.6]{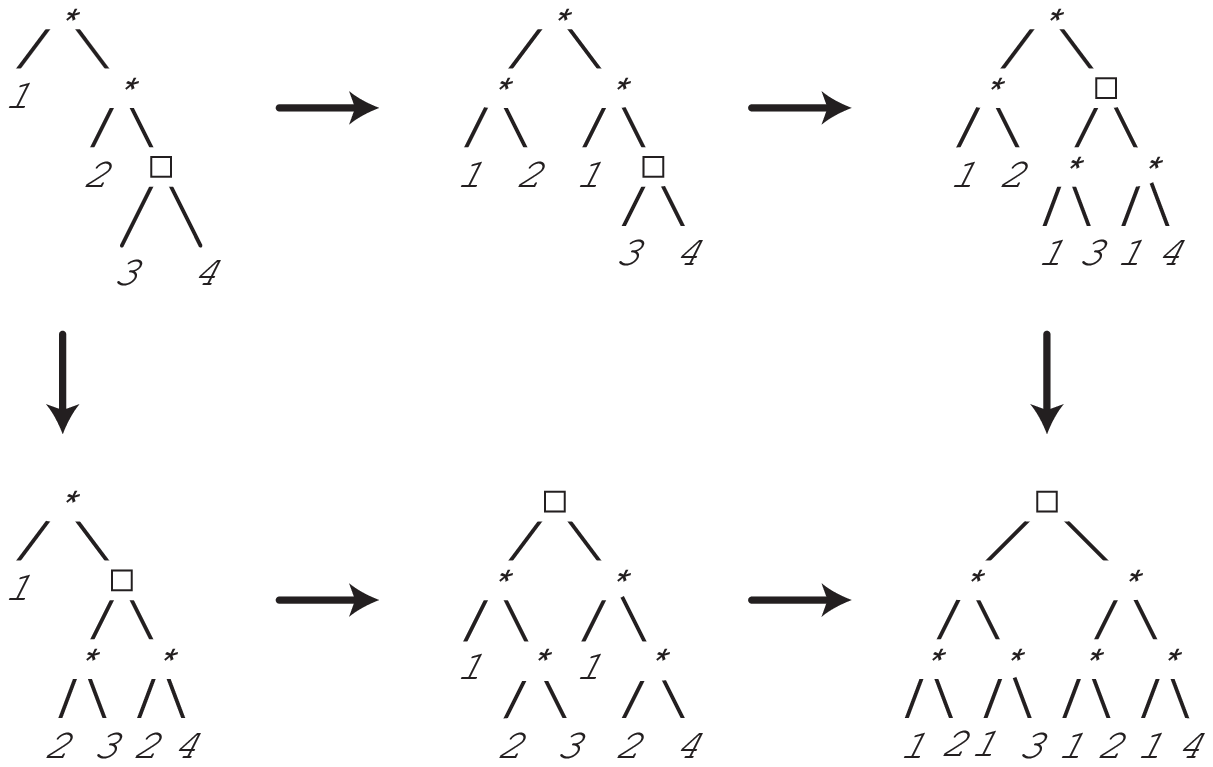}}
\put(7,23){$\SSp1$}
\put(67,23){$\SSp\ea$}
\put(19,43){$\SSp\ea$}
\put(48,43){$\SSp1$}
\put(19,13){$\SSp\ea$}
\put(45,13){$\SSp1\SSp0$}
\end{picture}

\begin{picture}(74,50)
\put(0,0){\includegraphics[scale=0.6]{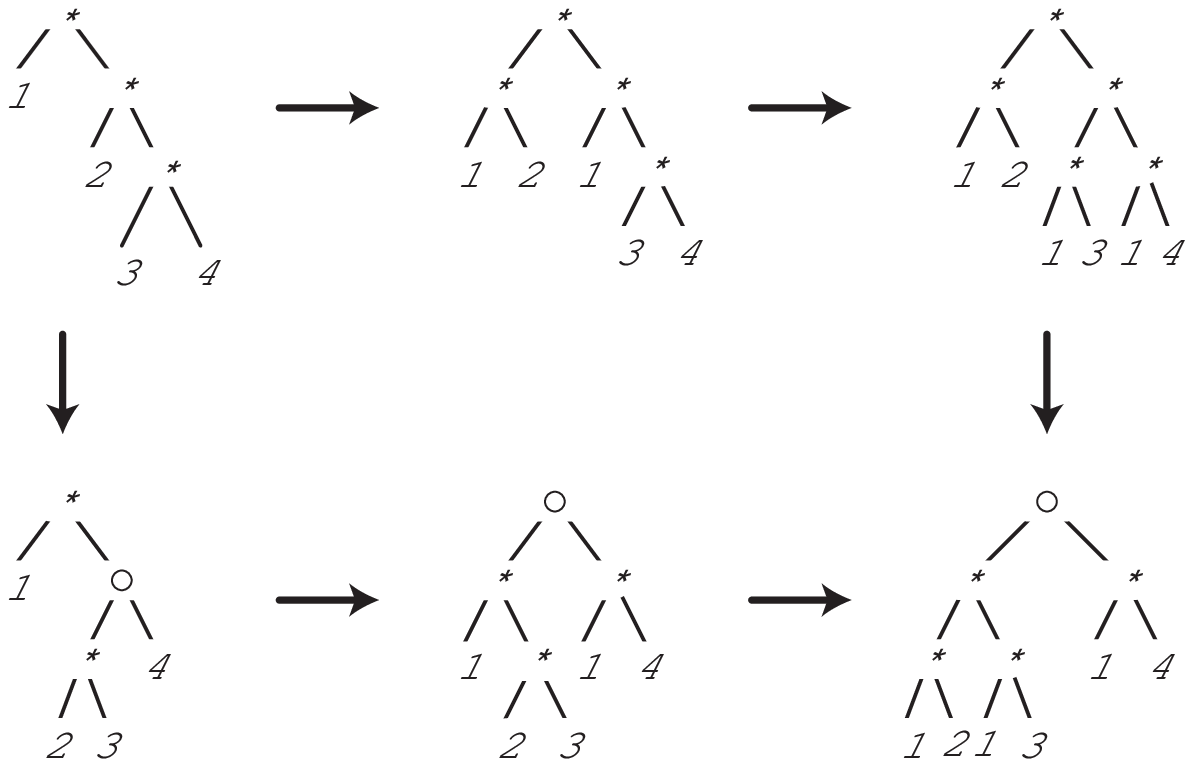}}
\put(7,23){$\AAp1$}
\put(67,23){$\AAp\ea$}
\put(19,43){$\SSp\ea$}
\put(48,43){$\SSp1$}
\put(19,13){$\SSp\ea$}
\put(48,13){$\SSp0$}
\end{picture}

\begin{picture}(74,50)
\put(0,0){\includegraphics[scale=0.6]{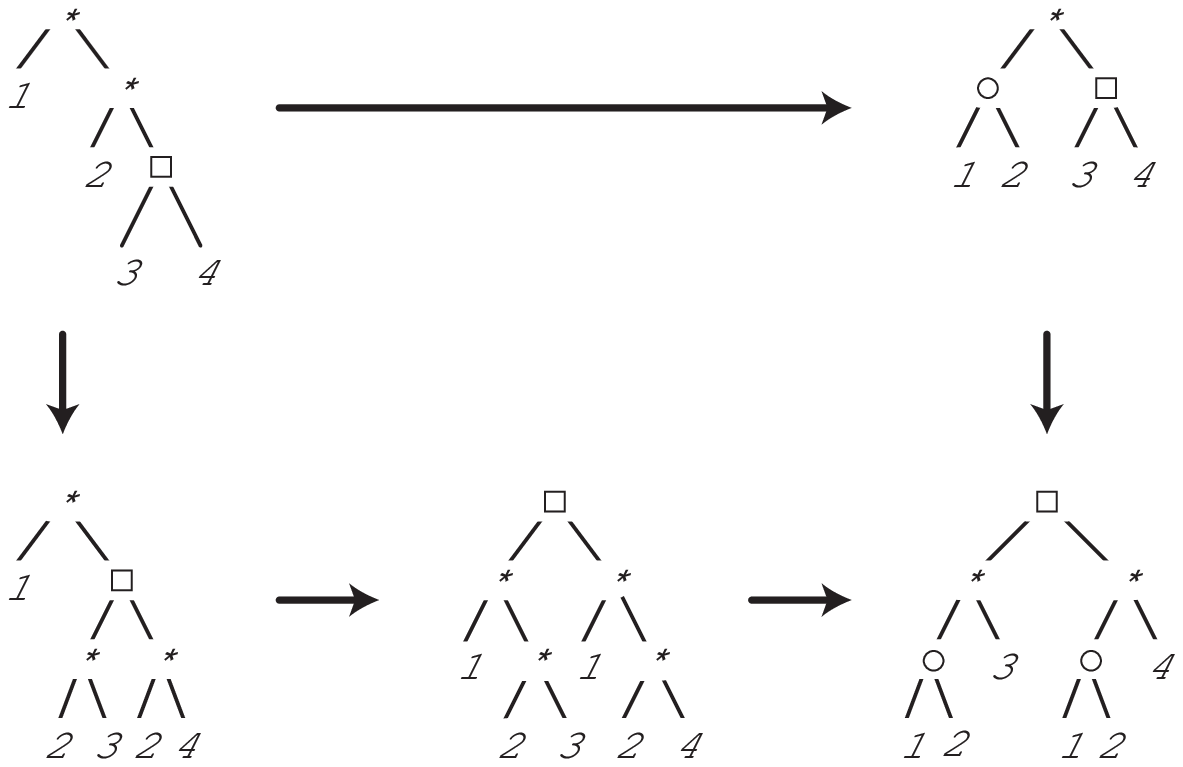}}
\put(7,23){$\SSp1$}
\put(67,23){$\SSp\ea$}
\put(34,43){$\AAp\ea$}
\put(19,13){$\SSp\ea$}
\put(45,13){$\AAp1\AAp0$}
\end{picture}
\caption{\smaller Three more types of confluence relations in $\GGALD$: here $\Op$ stands for both~$\op$ or~$\OP$, the numbers stand for the indices of the variables.}
\label{F:Crit}
\end{figure}

With the previous approach, we succeeded in finding one relation of type~\eqref{E:Conf} for all pairs~$\a, \b$, with the exception of the pairs $\{\AAp\a, \SSp\a\}$, as well as all pairs $\{\XXp\a, \YYp{\a1}\}$. So the question is whether we can find in~$\GGALDp$ relations of the form
\begin{equation*}
\XX\a \comp ... = \YY{\a} \comp ... 
\mbox{\quad and\quad}
\XX\a \comp ... = \YY{\a1} \comp ... 
\end{equation*}
when $(\XX{}, \YY{})$ ranges over the various combinations of~$\SS{}$ and~$\AA{}$. As shown in Figure~\ref{F:Crit}, there exist such relations, namely
\begin{gather}
\label{E:ALDCrit1}
\SSp\a \comp \SSp{\a1} \comp \SSp\a = 
\SSp{\a1} \comp \SSp\a \comp \SSp{\a1} \comp \SSp{\a0}, \\
\label{E:ALDCrit2}
\SSp\a \comp \SSp{\a1} \comp \AAp\a 
= \AAp{\a1} \comp \SSp\a \comp
\SS{\a0},\\
\label{E:ALDCrit3}
\AAp{\a} \comp \SSp{\a} = \SSp{\a1} \comp \SSp{\a} \comp \AAp{\a1} \comp \AAp{\a0}.
\end{gather}

Notice that all cases are not covered: there is {\it no} relation  $\AAp{\a} ... = \SSp{\a} ...$, 
or $\AAp{\a} ... = \AAp{\a1} ...$ in the list above.

\subsection{The geometry group}

According to the principle considered above, we introduce the abstract group for which the confluence relations of~$\GGLL$ make a presentation.

\begin{defi}
\label{D:GLL}
For~$\LLL$ a family of oriented algebraic laws, we define the {\it geometry group}~$\GLL$ of~$\LLL$ to be the group generated by formal copies of the operators~$\OOp\a$ for~$\LL$ in~$\LLL$, subject to all confluence relations connecting these operators in the monoid~$\GGLLp$. We denote by~$\ev$ the canonical evaluation morphism of~$\WLL$ onto~$\GLL$.
\end{defi}

\begin{exam}
In the case of the $\ALD$-laws, the group $\GALD$ is, by definition, a group generated by two infinite series of generators $\SS\a$ and $\AA\a$ indexed by finite sequences of~$0$'s and~$1$'s, subject to the relations~\eqref{E:ALDComm}, \eqref{E:ALDHeir1}, \eqref{E:ALDHeir2}, \eqref{E:ALDHeir3}, \eqref{E:ALDCrit1}, \eqref{E:ALDCrit2}, and \eqref{E:ALDCrit3}.
\end{exam}

\begin{exam}
Let us come back to the (linear) associativity law~$\Ass$. One easily checks that, for each pair of addresses $\a, \b$, there exists a natural confluence relation of the form $\AAp\a \comp ... = \AAp\b \comp ...$ in the positive monoid~$\GGAp$. Indeed, besides the quasi-commutativity inheritance relations
$$\AAp\a \comp \AAp{\a0\d} = \AAp{\a00\d} \comp \AAp\a, \quad
\AAp\a \comp \AAp{\a10\d} = \AAp{\a01\d} \comp \AAp\a, \quad
\AAp\a \comp \AAp{\a11\d} = \AAp{\a1\d} \comp \AAp\a,$$
the only missing pairs are the pairs $\a, \a1$, and the famous MacLane--Stasheff pentagon equality~\ref{Mac}
$$\AAp\a \comp \AAp\a = \AAp{\a1} \comp \AAp\a \comp \AAp{\a1}$$
completes the picture. Then the group~$\GA$ presented by the above relations coincides with the group $\GGA/\!\!\approx$ of Proposition~\ref{P:Geom}, \ie, with R.\,Thompson's group~$F$ \cite{Dfg}: so, in this case, the method based on presentation by confluence relations subsumes that of Section~\ref{SS:Linear}.
\end{exam}

When confluence relations are used to introduce the group~$\GLL$, the connection between the monoid~$\GGLL$ and the group~$\GLL$ is not obvious. Indeed, instead of the sequence of~\eqref{E:Quotient}, the scheme is now
\begin{equation}
\label{E:QuotientBis}
\begin{CD}
\WLL
@>{\mathrm{onto}}>{\eev}>
\GGLL\\
@V{\mathrm{onto}}V{\ev}V\\
\GLL
\end{CD}
\end{equation}
and it is not clear whether any factorization connects~$\GGLL$ and~$\GLL$. In order to prove that there exists a morphism from~$\GGLL$ to~$\GLL$, we ought to know that the selected confluence relations exhaust all relations holding in~$\GGLL$---which cannot be the case if the empty operator occurs. And, in order to prove that there exists a morphism from~$\GLL$ to~$\GGLL$, we ought to be able to control the pairs $\OOOp\LL\a \comp \OOOm\LL\a$ in~$\GGLL$. Precisely, assume that $\ww, \ww'$ are words in~$\WLL$ that satisfy $\ev(\ww) = \ev(\ww')$, \ie, that represent the same element of~$\GLL$. This means that there exists a finite sequence of words~$\ww = \ww_0, \ww_1, ..., \ww_n = \ww'$ such that each~$\ww_i$ is obtained from the previous one either by applying one of the confluence relation, or by deleting some subfactor $\OOO\LL\a \OOO\LL\a\inv$ or $\OOO\LL\a\inv \OOO\LL\a$, or by inserting such a subfactor. In the first two cases, the associated operators are equal, but, in the third case, the associated operators need not be equal, and, in particular, the empty operator may appear. So we cannot deduce from the fact that $\ww, \ww'$ represent the same element in~$\GLL$ the fact that they represent the same element in~$\GGLL$. However, and this is an important point, even if we cannot directly compare~$\GGLL$ and~$\GLL$, we can use 
the properties of~$\GGLL$ as a sort of oracle for guessing properties of~$\GLL$, and then try to find a direct proof for the results so conjectured.

\section{Internalization of terms}
\label{S:Blue}

At this point, we have have associated a monoid~$\GGLL$, and in some cases a group~$\GLL$, with each family of algebraic laws~$\LLL$. By construction, there is a partial action of the monoid~$\GGLL$ on terms, and similarly, at least in the case of linear laws, a partial action of the group~$\GLL$ on terms. The next step consists in trying to carry terms inside the monoid~$\GGLL$ and/or the group~$\GLL$, so as to replace the external action of the monoid or the group on terms by an internal operation, typically a multiplication in good cases. Once again, we have no general solution, but we explain how to complete the construction in good cases, and specifically in the case of the $\ALD$-laws.

\subsection{Blueprint of a term: the principle}
\label{SS:Blue1}

A priori, terms and operators of the geometry monoid live in disjoint worlds: the only connection is that operators act on terms. The principle we shall apply in the sequel---and which turns out to be efficient in good cases---consists in building inside the geometry monoid~$\GGLL$, and simultaneously inside the geometry group~$\GLL$ that mimicks its algebraic properties, a copy of each term, so that the action of operators on terms translates into a simple operation inside~$\GGLL$ and~$\GLL$, typically a multiplication. This copy of a term~$\tt$ in~$\GLL$ will be called the {\it blueprint} of~$\tt$.

We recall that, for~$\Sign$ a list of operation symbols, $\TS$ denotes the family of all terms constructed using the operations of~$\Sign$ and variables from an infinite list~$\Var$. In the sequel, we assume that $\xx$ is a fixed element of~$\Var$, and use $\TSx$ for the family of all terms constructined using the operations of~$\Sign$ and the single variable~$\xx$.

The general principle is as follows. Assume that $\LLL$ is a family of algebraic laws involving the signature~$\Sign$, and we have an injective map (``carbon copy'')
\begin{equation}
\IInt: \TSx \to \GGLL,
\end{equation}
\ie, a representation of terms inside the geometry monoid~$\GGLL$. If $\tt, \ttt$ are $\LLL$-equivalent terms, \ie, by Proposition~\ref{P:Basic}, if some element~$\ff$ of~$\GGLL$ maps~$\tt$ to~$\ttt$, there must exist some relation between the copies~$\IInt(\tt)$ and~$\IInt(\ttt)$. We shall be interested in the case when this relation takes the form
\begin{equation}
\label{E:Blue1}
\IInt(\tt \act \ff) \sim \IInt(\tt) \comp \varphi(\ff)
\end{equation}
where $\varphi$ is some endomorphism of~$\GGLL$, \ie, when the action of~$\GGLL$ on terms becomes a right multiplication twisted by endomorphism at the level of the copies---optimally, we might hope for an equality in~\eqref{E:Blue1}, but the problem of the empty operator makes a true equality impossible in most cases: this is precisely why we shall subsequently resort to the group~$\GLL$. We recall that $\WLL$ denotes the family of all abstract words in the letters~$\OOO\LL\a^{\pm1}$ for~$\LL$ in~$\LLL$ and~$\a$ an address. By construction, every operator in~$\GGLL$ is a finite product of elementary operators~$\OOO\LL\a^{\pm}$, so it can be expressed as $\eev(\ww)$ where $\ww$ is a word in~$\WLL$, and \eqref{E:Blue1} can be restated as
\begin{equation}
\label{E:Blue2}
\IInt(\tt \act \eev(\ww)) \sim \IInt(\tt) \comp \varphi(\eev(\ww))),
\end{equation}
for~$\tt$ in~$\TSx$ and $\ww$ in~$\WLL$ such that $\tt \act \eev(\ww)$ is defined.

Now, according to the general principle of Section~\ref{S:Group}, we wish to replace the monoid~$\GGLL$ with the group~$\GLL$. If the construction of the mapping~$\IInt$ is explicit enough, we can mimick it in the group~$\GLL$, thus defining a similar (hopefully injective) mapping
\begin{equation}
\Int: \TSx \to \GLL
\end{equation}
and use $\Int(\tt)$ as a copy of~$\tt$ inside~$\GLL$. If $\GLL$ resembles~$\GGLL$ enough, we may hope that the counterpart of~\eqref{E:Blue2} holds in~$\GLL$, so that we can carry the action of~$\GGLL$ on terms inside the group~$\GLL$. We are thus led to the following notion:

\begin{defi}
\label{D:Blue}
Assume that $\LLL$ is a family of oriented algebraic laws involving the signature~$\Sign$, and that $\phi$ is an endomorphism of~$\GLL$. A mapping $\Int: \TSx \to \GLL$ is said to be a {\it $\phi$-blueprint} if the relation
\begin{equation}
\label{E:Constr}
\Int(\tt \act \eev(\ww)) = \Int(\tt) \cdot \phi(\ev(\ww))
\end{equation}
holds in~$\GLL$ for every term~$\tt$ and every word~$\ww$ in~$\WLL$ such that $\tt \act \eev(\ww)$ is defined. 
\end{defi}

Thus, a $\phi$-blueprint transforms the operation of applying the laws of~$\LLL$ into a multiplication twisted by~$\phi$ in the group~$\GLL$. The interest of a $\phi$-blueprint will be discussed in Section~\ref{S:Using} below. The general idea is that it allows for carrying the problems about $\LLL$-equivalence inside the presented group~$\GLL$, and therefore may lead to solutions when the latter is under control. Typically, carrying the relation~$\EQ\LLL$ to~$\GLL$ using the map~$\Int$ supposed to be injective yields a new equivalence relation inside~$\GLL$ that may be more easily controlled than~$\EQ\LLL$ itself. 

\begin{rema}
A constant mapping provides a blueprint, certainly a trivial and uninteresting one. In the sequel, we are mainly interested in blueprints that are injective or close to, but we shall see that, in certain cases like that of~$\ALD$, even a blueprint that is not proved to be injective may be useful. That is why we do not require injectivity in Definition~\ref{D:Blue}.
\end{rema}

\subsection{Blueprint of a term: the example of~$\ALD$}

In order to realize the approach sketched above, \ie, to construct a blueprint, in the case of the~$\ALD$-laws, the first step consists in representing terms in~$\GGLL$, \ie, in selecting for each term~$\tt$ a certain operator~$\IInt(\tt)$ in the geometry monoid~$\GGLL$ that in some sense characterizes~$\tt$. A general idea is to choose an operator that {\it constructs}~$\tt$ starting from some fixed absolute startpoint. This procedure heavily depends on the specific algebraic laws we are investigating, here~$\ALD$. We use $\Tox$ for the family of all well-formed terms involving two binary operators~$\op, \OP$ and the single variable~$\xx$.

The solution we develop relies on some specific property of the $\ALD$-laws, namely the existence of an {\it absorption} phenomenon. To decribe this phenomenon, let us define {\it right vines} to be those terms of~$\Tox$ inductively specificed by
\begin{equation} \label{E:Vine}
\xx^{[n]} := \begin{cases}
\xx&\mbox{for $n = 1$}, \\
\xx \op \xx^{[n-1]}
&\mbox{for $n \ge 2$}.
\end{cases}
\end{equation}
The associated trees are ``all to the right'' trees. The following result expresses that, in presence of the laws of~$\ALD$, every term~$\tt$ of~$\Tox$ is absorbed by all sufficiently large right vines.

\begin{lemm} [absorption lemma]
\label{L:Absorption}
For each term~$\tt$ in~$\Tox$, there exists a positive integer~$p$ such that 
\begin{equation} \label{E:Absorption}
\xx^{[n]} \EQ\ALD \tt \op \xx^{[n-p]}
\mbox{\qquad holds for $n$ large enough}.
\end{equation}
\end{lemm}

\begin{proof}
The property is true with $p = 1$ for~$\tt = \xx$ and $n \ge 2$ (in which case the equivalence is an equality), so, in order to establish it for every term in~$\Tox$, it
is enough to show that, if
\eqref{E:Absorption} holds for~$\tti$
and~$\ttii$, then it holds for
$\tti \op \ttii$ and for $\tti \OP \ttii$ as
well. So assume $\xx^{[n]} \EQ\ALD \tti \op
\xx^{[n-p_1]}$ for $n \ge m_1$ and
$\xx^{[n]} \EQ\ALD \ttii
\op \xx^{[n-p_2]}$ for $n \ge m_2$. We
obtain for $n \ge
\max(m_1+p_2, m_2+p_1)$
\begin{align*}
\xx^{[n]}
&\EQ\ALD \tti \op \xx^{[n-p_1]} 
&\mbox{by hypothesis,}\\
&\EQ\ALD \tti \op (\ttii \op \xx^{[n-p_1 -p_2]}) 
&\mbox{by hypothesis,}\\
&\EQ\ALD (\tti \op \ttii) \op (\tti \op \xx^{[n-p_1 -p_2]}) 
&\mbox{by $\LD$,}\\
&\EQ\ALD (\tti \op \ttii) \op \xx^{[n-p_2]} 
&\mbox{by hypothesis.}
\end{align*}
Similarly, we have for $n \ge \max(m_1,
m_2+p_1)$
\begin{align*}
\xx^{[n]}
&\EQ\ALD \tti \op \xx^{[n-p_1]} 
&\mbox{by hypothesis,}\\
&\EQ\ALD \tti \op (\ttii \op \xx^{[n-p_1 -p_2]}) 
&\mbox{by hypothesis,}\\
&\EQ\ALD (\tti \OP \ttii) \op \xx^{[n-p_1 -p_2]}
&\mbox{by $\ALDii$,}
\end{align*}
so the induction is completed.
\end{proof}

By Proposition~\ref{P:Basic}, the equivalence of~\eqref{E:Absorption} must be witnessed for by some operator of the geometry monoid~$\GGALD$: for each term~$\tt$ in~$\Tox$ and every integer~$n$, there must exist an operator~$\IInt(\tt)$ in~$\GGALD$, hence a composition of operators~$\SS\a^\pm$ and~$\AA\a^\pm$, that maps~$\xx^{[n]}$ to $\tt \op \xx^{[n-p]}$. Actually, the inductive proof of Lemma~\ref{L:Absorption} gives not only the existence of such a witness, but only an inductive construction for such a witness.

\begin{lemm} \label{L:Translation}
For~$\tt$ in~$\Tox$, inductively define $\IInt(\tt)$ in~$\GGALD$ by
\begin{equation}
\IInt(\tt) = 
\begin{cases}
\id 
& \mbox{for $\tt = \xx$,}\\
\IInt(\tti) \comp \ssh1(\IInt(\ttii)) \comp \SSp\ea \comp \ssh1(\IInt(\tti))\inv 
&\mbox{for $\tt = \tti \op \ttii$,}\\
\IInt(\tti) \comp \ssh1(\IInt(\ttii)) \comp \AAp\ea  
&\mbox{for $\tt = \tti \OP \ttii$}.
\end{cases}
\end{equation}
Then, for every term~$\tt$ in~$\Tox$, there exists~$p$ such that, for every~$n$ large enough, we have
\begin{equation}
\label{E:Blue}
\IInt(\tt) : \xx^{[n]} \mapsto \tt \op \xx^{[n-p]}.
\end{equation}
\end{lemm}

\begin{proof}
The formulas of~\eqref{E:NNewop} are a mere translation of the successive equivalence steps in the proof of Lemma~\ref{L:Absorption}, and the result is then a straightforward verification.
\end{proof}

Note that \eqref{E:Blue} guarantees that the mapping~$\IInt$ is injective, since the term~$\tt$ can be recovered from the operator~$\IInt(\tt)$. So it is coherent to use the operator~$\IInt(\tt)$ as a counterpart of the term~$\tt$ inside the geometry monoid~$\GGALD$. According to the general scheme of Section~\ref{SS:Blue1}, we shall now analyse the counterpart of the action of~$\GGLL$ on terms. 

\begin{lemm}
\label{L:Action}
For each term~$\tt$ in~$\Tox$ and each operator~$\ff$ in~$\GGALD$ such that $\tt \act \ff$ is defined, we have
\begin{equation}
\label{E:AAction}
\IInt(\tt \act \ff) \sim \IInt(\tt) \comp \ssh0(\ff).
\end{equation}
\end{lemm}

\begin{proof}
Let $\ttt = \tt \act \ff$. Then, for each term~$\tti$, the operator~$\sh0(\ff)$ maps the term~$\tt \op \tti$ to~$\ttt \op \tti$---and, similarly, $\tt \OP \tti$ to $(\tt \act \ff) \OP \tti$. So, in particular, $\sh0(\ff)$ maps~$\tt \op \xx^{[n]}$ to~$\ttt \op \xx^{[n]}$ for each~$n$. Now, by Lemma~\ref{L:Translation} (and for~$n$ large enough), the operator~$\IInt(\tt)$ maps~$\xx^{[n]}$ to~$\tt \op \xx^{[n-p]}$, while $\IInt(\ttt)$ maps~$\xx^{[n]}$ to~$\ttt \op \xx^{[n - p']}$ for some~$p'$. This means that both $\IInt(\tt) \comp \sh0(\ff)$ and $\IInt(\ttt)$ map~$\xx^{[n]}$ to~$\ttt \op \xx^{[n-p']}$. Hence, in the monoid~$\GGALD$, the two operators $\IInt(\tt) \comp \sh0(\ff)$ and $\IInt(\ttt)$ agree on at least one term, namely~$\tt$, which, by definition, means that \eqref{E:AAction} holds.
\end{proof}

We thus obtained in the case of~$\ALD$ a relation of the form~\eqref{E:Blue1}, the involved endormorphism of~$\GGALD$ being~$\ssh0$. Expressed on the shape of~\eqref{E:Blue2}, the relation reads
\begin{equation}
\label{E:AActionBis}
\IInt(\tt \act \eev(\ww)) \sim \IInt(\tt) \comp \ssh0(\eev(\ww))
\end{equation}
for $\tt$ a term in~$\Tox$ and $\ww$ a word in~$\WALD$ such that $\tt \act \eev(\ww)$ is defined. 

Following the general scheme of Section~\ref{SS:Blue1} again, we now mimick the construction of~$\IInt$ inside the group~$\GALD$.  Using $\sh\a$ to denote the endomorphism of~$\GALD$ that maps $\SS\a$ to~$\SS{1\a}$ and $\AA\a$ to~$\AA{1\a}$ for each address~$\a$, this amounts to setting:

\begin{defi}
We inductively associate with every term~$\tt$ in~$\Tox$ an element~$\Int(\tt)$ of~$\GALD$, also denoted~$\blue\tt$, by
\begin{equation}
\Int(\tt) = 
\begin{cases}
1 
&\mbox{for $\tt = \xx$,}\\
\Int(\tti) \cdot \sh1(\Int(\ttii)) \cdot \SS\ea \cdot \sh1(\Int(\tti))\inv 
&\mbox{for $\tt = \tti \op \ttii$,}\\
\Int(\tti) \cdot \sh1(\Int(\ttii)) \cdot \AA\ea  
&\mbox{for $\tt = \tti \OP \ttii$.}
\end{cases}
\end{equation}
\end{defi}

\begin{exam}
Let $\tt$ be $\xx \op ((\xx \OP \xx) \op \xx)$. Starting from $\Int(\xx) = 1$, we first find 
$$\Int(\xx \OP \xx) = 1 \cdot \sh1(1) \cdot \AA\ea = \AA\ea,$$ 
then
$$\Int((\xx \OP \xx) \op \xx) = \AA\ea \cdot \sh1(1) \cdot \SS\ea \cdot \sh1(\AA\ea)\inv = \AA\ea \SS\ea \AA1\inv,$$ 
and, finally,
$$\Int(\tt) = 1 \cdot \sh1(\AA\ea \SS\ea \AA1\inv) \cdot \SS\ea \cdot \sh1(1)\inv = \AA1 \SS1 \AA{11}\inv \SS\ea.$$ 
\end{exam}

If our intuition is correct, \ie, if the confluence relations defining the group~$\GALD$ capture enough of the geometry of the laws~$\ALD$, Relation~\eqref{E:AActionBis} should follow from the lattice relations in~$\GGALD$, and, therefore, it should induce an equality in the group~$\GALD$, \ie, we {\it should} have the relation
\begin{equation}
\label{E:ActionBis}
\Int(\tt \act \eev(\ww)) = \Int(\tt) \cdot \sh0(\ev(\ww)),
\end{equation}
\ie, with our former definition, the mapping~$\Int$ should be an $\sh0$-blueprint . This is indeed the case---and this is the key point for our current analysis of the ALD-laws. The nice feature is that, if the result is true---and it is---its proof must be a simple verification.

\begin{lemm}
\label{L:Main}
The mapping~$\Int$ is a $\sh0$-blueprint for the $\ALD$-laws.
\end{lemm}

\begin{proof}
For an induction on the length of the word~$\ww$, it is enough to prove the result when $\ww$ consists of a single letter, \ie, it is one of $\AA\a^{\pm 1}$, $\SS\a^{\pm1}$. Moreover, the cases of $\AA\a\inv$ and $\SS\a\inv$ immediately follow from the cases of~$\AA\a$ and~$\SS\a$, respectively. So, the point is to establish the equalities
\begin{equation}
\label{E:ActionLocal}
\Int(\tt \act \AA\a) = \Int(\tt) \cdot \AA{0\a} ,\quad
\Int(\tt \act \SS\a) = \Int(\tt) \cdot \SS{0\a},
\end{equation}
whenever the involved terms are defined. 

We prove~\eqref{E:ActionLocal} using induction on the length of the address~$\a$. Let us begin with $\a = \ea$ and the case of~$\AA\a$, \ie, of~$\AA\ea$. Saying that $\tt \act \AA\ea$ (\ie, $\tt \act \AA\ea$) is defined, means that $\tt$ can be decomposed as $\tt = \tti \op (\ttii \op \ttiii)$, and, then, we have $\ttt = (\tti \OP \ttii) \op \ttiii$. Using the commutation and quasi-commutation relations of~$\GALD$, we find
\begin{align*}
\blue\ttt 
&= \blue\tti \cdot \sh1(\blue\ttii) \cdot \AA\ea \cdot \sh1(\blue\ttiii) \cdot \SS\ea \cdot \AA1\inv \cdot \sh{11}(\blue\ttii)\inv \cdot \sh1(\blue\tti)\inv ,\\
&= \blue\tti \cdot \sh1(\blue\ttii) \cdot \sh{11}(\blue\ttiii) \cdot \AA\ea \cdot \SS\ea \cdot \AA1\inv \cdot \sh{11}(\blue\ttii)\inv \cdot \sh1(\blue\tti)\inv ,\\
\blue\tt \cdot \AA{0\a} 
&= \blue\tti \cdot \sh1(\blue\ttii) \cdot \sh{11}(\blue\ttiii) \cdot \SS1 \cdot \sh{11}(\blue\ttii)\inv \cdot \SS\ea \cdot \sh1(\blue\tti)\inv \cdot \AA0,\\
&= \blue\tti \cdot \sh1(\blue\ttii) \cdot \sh{11}(\blue\ttiii) \cdot \SS1 \cdot \SS\ea \cdot \AA0 \cdot \sh{11}(\blue\ttii)\inv \cdot \sh1(\blue\tti)\inv,
\end{align*}
and the equality follows from~\eqref{E:ALDCrit3}, which gives 
$\AA\ea \cdot \SS\ea \cdot \AA1\inv = \SS1 \cdot \SS\ea \cdot \AA0$.

We consider now the case of~$\SS\ea$. The hypothesis that $\tt \act \SS\ea$ is defined means that $\tt$ can be decomposed as $\tti \op (\ttii \Op \ttiii)$, and, then, $\ttt$ is $(\tti \op \ttii) \Op (\tti \op \ttiii)$. Assume first $\Op = \op$ (this is the most complicated case). Using the commutation and quasi-commutation relations of~$\GALD$, we find
\begin{align*}
\blue\ttt 
&= \blue\tti \cdot \sh1(\blue\ttii) \cdot \SS\ea \cdot \sh1(\blue\tti)\inv \cdot \sh1(\blue\tti) \cdot \sh{11}(\blue\ttiii) \cdot \SS1 \cdot \sh{11}(\blue\tti)\inv \\
&\hspace{40mm}
\cdot \SS\ea \cdot \sh{11}(\blue\tti) \cdot \SS1\inv \cdot \sh{11}(\blue\ttii)\inv \cdot \sh1(\blue\tti)\inv\\
&= \blue\tti \cdot \sh1(\blue\ttii)  \cdot \sh{11}(\blue\ttiii) \cdot \SS\ea \cdot \SS1 
\cdot \SS\ea \cdot \SS1\inv \cdot \sh{11}(\blue\ttii)\inv \cdot \sh1(\blue\tti)\inv,\\
\blue\tt \cdot \SS{0\a} 
&= \blue\tti \cdot \sh1(\blue\ttii) \cdot \sh{11}(\blue\ttiii) \cdot \SS1 \cdot \sh{11}(\blue\ttii)\inv \cdot \SS\ea \cdot \sh1(\blue\tti)\inv \cdot \SS0\\
&= \blue\tti \cdot \sh1(\blue\ttii) \cdot \sh{11}(\blue\ttiii) \cdot \SS1 \cdot \SS\ea \cdot \SS0 \cdot \sh{11}(\blue\ttii)\inv \cdot \sh1(\blue\tti)\inv,
\end{align*}
and the equality follows from the~\eqref{E:ALDCrit1} relation $\SS1 \cdot \SS\ea \cdot \SS1 \cdot \SS0 = \SS\ea \cdot \SS1 \cdot \SS\ea$. Assume now $\Op = \OP$. One finds
\begin{align*}
\blue\ttt 
&= \blue\tti \cdot \sh1(\blue\ttii) \cdot \SS\ea \cdot \sh1(\blue\tti)\inv \cdot \sh1(\blue\tti) \cdot \sh{11}(\blue\ttiii) \cdot \SS1 \cdot \sh{11}(\blue\tti)\inv \cdot \AA\ea \\
&= \blue\tti \cdot \sh1(\blue\ttii) \cdot \sh{11}(\blue\ttiii) \cdot \SS\ea \cdot \SS1 \cdot \AA\ea\cdot \sh1(\blue\tti)\inv ,\\
\blue\tt \cdot \SS{0\a} 
&= \blue\tti \cdot \sh1(\blue\ttii) \cdot \sh{11}(\blue\ttiii) \cdot \AA1 \cdot \SS\ea \cdot \sh1(\blue\tti)\inv \cdot \SS0\\
&= \blue\tti \cdot \sh1(\blue\ttii) \cdot \sh{11}(\blue\ttiii) \cdot \AA1 \cdot \SS\ea  \cdot \SS0\cdot \sh1(\blue\tti)\inv,
\end{align*}
and the equality follows from the~\eqref{E:ALDCrit2} relation $\AA1 \cdot \SS\ea \cdot \SS0 = \SS\ea \cdot \SS1 \cdot \AA\ea$. 

The case $\a = \ea$ is completed. 
From now on, we shall treat the cases of~$\AA\a$ and~$\SS\a$ simultaneously, using~$\XX\a$ as a common notation. Assume first that the address~$\a$ is $0\b$ for some~$\b$. The hypothesis that $\tt \act \XX\a$ is defined implies that $\tt$ can be decomposed as $\tti \Op \ttii$, and, then, $\ttt$ is $\ttti \Op \ttii$, with $\ttti = \tti \act \XX\b$. Assume first $\Op = \op$. The induction hypothesis implies $\blue\ttti = \blue\tti \cdot \XX{0\b}$. We find
\begin{align*}
\blue\ttt 
&= \blue\tti \cdot \XX{0\b} \cdot \sh1(\blue\ttii) \cdot \SS\ea \cdot \XX{10\b}\inv \cdot \sh1(\blue\tti)\inv ,\\
&= \blue\tti \cdot \sh1(\blue\ttii) \cdot \XX{0\b} \cdot \SS\ea \cdot \XX{10\b}\inv \cdot \sh1(\blue\tti)\inv,\\
\blue\tt \cdot \XX{0\a} 
&= \blue\tti \cdot \sh1(\blue\ttii) \cdot \SS\ea \cdot \sh1(\blue\tti)\inv \cdot \XX{00\b},\\
&= \blue\tti \cdot \sh1(\blue\ttii) \cdot \SS\ea \cdot \XX{00\b} \cdot \sh1(\blue\tti)\inv,
\end{align*}
and the equality follows from the~\eqref{E:ALDHeir1} relation $\XX{0\b} \cdot \SS\ea = \SS\ea \cdot \XX{00\b} \cdot \XX{10\b}$. Similarly, for $\Op = \OP$, we find
\begin{align*}
\blue\ttt 
&= \blue\tti \cdot \XX{0\b} \cdot \sh1(\blue\ttii) \cdot \AA\ea 
= \blue\tti \cdot \sh1(\blue\ttii) \cdot \XX{0\b} \cdot \AA\ea ,\\
\blue\tt \cdot \XX{0\a} 
&= \blue\tti \cdot \sh1(\blue\ttii) \cdot \AA\ea \cdot \XX{00\b},
\end{align*}
and the equality follows from the~\eqref{E:ALDHeir4} relation $\XX{0\b} \cdot \AA\ea = \AA\ea \cdot \XX{00\b}$. 

The argument is similar when  the address~$\a$ is $1\b$ for some~$\b$. The hypothesis that $\tt \act \XX\a$ is defined implies that $\tt$ can de decomposed as $\tti \Op \ttii$, and, then, $\ttt$ is $\tti \Op \tttii$, with $\tttii = \ttii \act \XX\b$. Assume first $\Op = \op$. The induction hypothesis implies $\blue\ttti = \blue\tti \cdot \XX{0\b}$. We find now
\begin{align*}
\blue\tt \cdot \XX{0\a} 
&= \blue\tti \cdot \sh1(\blue\ttii) \cdot \SS\ea \cdot \sh1(\blue\tti)\inv \cdot \XX{01\b}
= \blue\tti \cdot \sh1(\blue\ttii) \cdot \SS\ea \cdot \XX{01\b} \cdot \sh1(\blue\tti)\inv,\\
\blue\ttt 
&= \blue\tti \cdot \sh1(\blue\ttii) \cdot \XX{10\b} \cdot \SS\ea \cdot \sh1(\blue\tti)\inv ,
\end{align*}
and the equality follows from the~\eqref{E:ALDHeir2} relation $\XX{10\b} \cdot \SS\ea = \SS\ea \cdot \XX{01\b}$. Finally, for $\Op = \OP$, we have
\begin{align*}
\blue\tt \cdot \XX{0\a} 
&= \blue\tti \cdot \sh1(\blue\ttii) \cdot \AA\ea \cdot \XX{01\b},\\
\blue\ttt 
&= \blue\tti \cdot \sh1(\blue\ttii) \cdot \XX{01\b} \cdot \AA\ea,
\end{align*}
and the equality follows from the~\eqref{E:ALDHeir5} relation $\XX{10\b} \cdot \AA\ea = \AA\ea \cdot \XX{01\b}$. The induction is complete. 
\end{proof}

We thus completed the construction of a blueprint for the $\ALD$-laws. It may be observed that this construction induces the construction of a similar $\sh0$-blueprint for the $\LD$-law considered alone: indeed, in the case of terms not containing the operator~$\OP$, the blueprint does not involve any generator~$\AA\a$, and it can be checked that the only relations needed to check the blueprint condition are present in the group~$\GLD$.

\begin{rema}
Here the blueprints have been defined for terms in one variable only. Developing a similar approach for terms involving several variables is possible, at the expense of extending the geometry monoid~$\GGLL$ by introducing additional operators whose action is to shift the indices of the variables so as to still generate all terms starting from right vines. We refer to~\cite{Dgd} for details in the specific case of the $\LD$-law.
\end{rema}

\section{Using~$\GLL$ to study~$\LLL$}
\label{S:Using}

We claim that the geometry group~$\GLL$ is an interesting object that provides useful information about the laws of~$\LLL$. The way this vague statement can be made precise depends on the specific algebraic laws one considers. If the latter are simple, typically associativity and/or commutativity, constructing an $\LLL$-algebra or solving the word problem of~$\LLL$ is not a challenge, and in particular appealing to~$\GLL$ is not necessary. However the group~$\GLL$ may be of interest in itself, as the enormous literature devoted to R.Thompson's groups~$F$ and~$V$ shows~\cite{Tho, McT, CFP}. In the case of complicated laws, typically self-distributivity or variants, constructing $\LLL$-algebras and solving the word problem of~$\LLL$ may be difficult (and often even open) questions, and then the geometry group may be useful. The basic scheme consists in exploiting the blueprint construction---when it exists---to define an algebraic system satisfying the laws of~$\LLL$ on some quotient of the group~$\GLL$.

\subsection{Construction of algebraic systems: principle}

The first, and more direct, application of the previous approach is the construction of an algebra that obeys some prescribed laws. The general principle is as follows:

\begin{prop}
\label{P:Eval}
Assume that $\LLL$ is a family of balanced algebraic laws involving the signature~$\Sign$, that $\Int$ is a $\phi$-blueprint for the laws~$\LLL$, and that $H$ is a subgroup of~$\GLL$ that includes the image of~$\phi$. For~$\tt$ a term in~$\TSx$, let $\cl(\tt)$ be the left coset of~$\Int(\tt)$ modulo~$H$. Then the map~$\cl$ is constant on each $\EQ\LL$-class, \ie, $\LLL$-equivalent terms have the same image.
\end{prop}

\begin{proof}
Assume that $\tt$ and $\ttt$ are $\LLL$-equivalent terms in~$\Tox$. By Proposition~\ref{P:Basic}, there must exist an operator in~$\GGLL$ that maps~$\tt$ to~$\ttt$, and, therefore, there must exist a word~$\ww$ in~$\WLL$ such that $\tt \act \eev(\ww)$ is defined and equal to~$\ttt$. Then, by~\eqref{E:Constr}, we have $\Int(\ttt) = \Int(\tt) \cdot \phi(\ev(\ww))$ in~$\GLL$, so the hypothesis gives
$$(\Int(\tt))\inv \cdot (\Int(\ttt)) \in H,$$
\ie, $\Int(\tt)$ and $\Int(\ttt)$ lie in the same $H$-coset.
\end{proof}

\begin{coro}
\label{C:Eval}
Under the same hypothesis, let $M$ be the image of the mapping~$\cl$. For each operation symbol~$\Op$ in~$\Sign$, define an operation in~$M$ by 
$$\Int(\tt) H \Op \Int(\ttt) H := \Int(\tt \Op \ttt)H.$$
Then $M$ equipped with these operations is an $\LLL$-algebra, \ie, an $\Sign$-structure satisfying all laws of~$\LLL$. 
\end{coro}

Note that, in general, nothing guarantees that the system so obtained is non-trivial: it might happen that the quotient-structure $H\backslash\GLL$ collapses to a point. This does not happen in the cases we shall consider: on the contrary, the obtained algebraic systems will turn out to be free, \ie, as far from trivial as possible.

\subsection{Construction of algebraic systems: the example of~$\ALD$}

We return to our leading example, namely the $\ALD$-laws. In Section~\ref{S:Blue} we constructed a $\sh0$-blueprint~$\Int$ for~$\ALD$, so Proposition~\ref{P:Eval} and Corollary~\ref{C:Eval} directly apply for each subgroup~$H$ of~$\GALD$ including the image of~$\sh0$, typically $H := \sh0(\GALD)$. We thus obtain an $\ALD$-algebra whose domain is some subset of the coset set $\sh0(\GALD)\backslash\GALD$. 

Actually, we can adapt the results to make them more simple and handy. Indeed, instead of restricting to the image of the mapping~$\cl$, we can extend the construction to the whole group~$\GALD$. To do that, the idea is obvious: we look at the inductive definition of the blueprint, and define $\op$ and $\OP$ to be the operations used to construct $\Int(\tt \op \ttt)$ and $\Int(\tt \OP \ttt)$ from $\Int(\tt)$ and $\Int(\ttt)$, \ie, we choose the operations on~$\GALD$ that make $\Int$ a morphism from the free algebra~$(\Tox, \op, \OP)$ to~$\GALD$.

\begin{lemm}
\label{L:Obst}
On the group~$\GALD$ define two new binary operations $\op, \OP$ by
\begin{equation} 
\label{E:Newop}
\xx \op \yy := \xx \cdot \sh1(\yy) \cdot
\SS\ea \cdot \sh1(\xx)\inv, \qquad
\xx \OP \yy := \xx \cdot \sh1(\yy) \cdot
\AA\ea.
\end{equation}
Then, for all $\xx, \yy, \zz$ in~$\GALD$ and $\Op$ in $\{\op, \OP\}$, we have
\begin{gather}
\label{E:Obst1}
(\xx \op \yy) \Op (\xx \op \zz) = x \op (\yy \Op \zz) \cdot \SS0,\\
\label{E:Obst2}
(\xx \OP \yy) \op \zz = \xx \op (\yy \op \zz) \cdot \AA0,\\
\label{E:Quot1}
(\xx \cdot \sh0(\zz) \Op \yy = \xx \Op \yy \cdot \sh{00}(\zz),\\
\label{E:Quot2}
\xx \Op (\yy \cdot \sh0(\zz)) = x \Op \yy \cdot \sh{01}(\zz).
\end{gather}
\end{lemm}

\begin{proof}
The verifications are those already made in the proof of Lemma~\ref{L:Main}. The only difference is that, in Section~\ref{S:Blue}, we only consider elements of~$\GALD$ 
that are blueprints of terms, while, now, we consider arbitrary elements of~$\GALD$. Now inspecting the proof of Section~\ref{S:Blue} shows that the specific form of the elements is never used, and, therefore, the whole computation remains valid.
\end{proof}

Relations~\eqref{E:Obst1} and~\eqref{E:Obst2} control the obstruction for $(\GALD, \op, \OP)$ to be an $\ALD$-algebra, and show that the latter belongs to the subgroup~$\sh0(\GALD)$. Relations~\eqref{E:Quot1} and~\eqref{E:Quot2} show that the operations on~$\GALD$ induce well-defined operations on the coset set $\sh0(\GALD) \backslash \GALD$. Thus we may state:

\begin{prop}
Let $M$ be the coset set~$\sh0(\GALD) \backslash \GALD$. Then $M$ equipped with the operations induced by those of Lemma~\ref{L:Obst} is an $\ALD$-algebra.
\end{prop}

The subgroup~$\sh0(\GALD)$ is not normal, and, therefore, the associated coset set is not a group. When we replace $\sh0(\GALD)$ with a larger subgroup~$H$ of~$\GALD$, typically the normal subgroup generated by $\sh0(\GALD)$, we can still apply Proposition~\ref{P:Eval}, but it is not a priori sure that the operations~$\op$ and~$\OP$ induce well-defined operations on the whole of $H \backslash \GALD$. This however happens in good cases, as here with~$\ALD$.

\begin{lemm}
Every normal subgroup of~$\GALD$ that includes the image of~$\sh0$ contains all generators~$\SS\a$ and~$\AA\a$ such that $\a$ contains at least one~$0$.
\end{lemm}

\begin{proof}
For $\XX{} = \SS{}$ or~$\AA{}$, the commutation relation~\eqref{E:ALDHeir5} gives $\XX{10} = \AA\eaÊ\cdot \XX{01} \cdot \AA\ea\inv$, hence, inductively
$$\XX{1^i0\a} = \AA{1^{i-1}} \cdot \ldots \cdot \AA1 \cdot \AA\ea \cdot \XX{01^i\a} \cdot \AA\ea\inv \cdot \AA1\inv \cdot \ldots \cdot \AA{1^{i-1}}\inv,$$
which shows that ant normal subgroup containing all~$\XX{0\g}$ must contain all~$\XX\a$ such that $\a$ contains at least one~$0$.
\end{proof}

Thus, collapsing all generators~$\SS\a$ and~$\AA\a$ such that $\a$ begins with~$0$ in~$\GALD$ requires to collapse all generators~$\XX\a$ such that $\a$ contains at least one~$0$, in which case the quotient-group is generated by the images of the remaining generators, namely the generators~$\SS{1^{i-1}}$ and~$\AA{1^{i-1}}$ with $i \ge 1$. Considering what remains from the defining relations of~$\GALD$, and using~$\ss i$ and~$\aa i$ as simplified notation for~$\SS{1^{i-1}}$ and~$\AA{1^{i-1}}$, we are led to the following group:

\begin{defi}
\label{D:Braids}
We let $\Bp$ be the group generated two infinite sequences $\ss1, \ss2, ...$ and $\aa1, \aa2, ...$ of generators subject to the relations
\begin{equation} 
\label{E:Relations}
\begin{cases}
\ \ss i \ss j  = \ss j \ss i,
\quad \ss i \aa j  = \aa j \ss i,
\quad \aa i \aa{j-1} = \aa j \aa i,
\quad \aa i \ss{j-1} = \ss j \aa i,\\
\ \ss i \ss{i+1} \ss i 
= \ss{i+1} \ss i \ss{i+1},
\quad
\ss{i+1} \ss i \aa{i+1}
= \aa i \ss i,
\quad
\ss i \ss{i+1} \aa i 
= \aa{i+1} \ss i
\end{cases}
\end{equation}
for $i \ge 1$ and $j \ge i+2$.
\end{defi}

(We do not claim that $\Bp$ is the quotient of~$\GALD$ by the normal subgroup generated by~$\sh0(\GALD)$.) By construction, mapping $\SS{1^{i-1}}$ to~$\ss i$ and $\AA{1^{i-1}}$ to~$\aa i$ defines a surjective homomorphism~$\pi$ of~$\GALD$ onto~$\Bp$ whose kernel includes~$\sh0(\GALD)$, and therefore the normal subgroup~$N$ it generates. However, it might be that $\pi$ collapses more than~$N$, and proving that this does not happen would require a more complete algebraic study of the group~$\GALD$, which is not our current aim. Here, the only thing we wish to observe that, by mimicking once again the construction of the blueprint, we can construct an $\ALD$-structure on the group~$\Bp$:

\begin{prop}
\cite{Dhe, Dhj}
Let $\sh{}$ denote the endomorphism of the group~$\Bp$ that maps~$\ss i$ to~$\ss{i+1}$ and~$\aa i$ to~$\aa{i+1}$ for every positive integer~$i$. Define binary operations~$\op, \OP$ on~$\Bp$ by
\begin{equation}
\label{E:NNewop}
\xx \op \yy := \xx \cdot \sh{}(\yy) \cdot
\ss1 \cdot \sh1(\xx)\inv, \qquad
\xx \OP \yy := \xx \cdot \sh{}(\yy) \cdot \aa1.
\end{equation}
Then $(\Bp, \op, \OP)$ is an $\ALD$-algebra. Moreover, this $\ALD$-algebra is torsion-free, \ie, every element of~$\Bp$ generates a free $\ALD$-subsystem.
\end{prop}
 
The group~$\Bp$ was introduced by M.\,Brin in~\cite{Bri1, Bri2} and the author in~\cite{Dhb} independently, and its elements have been interpreted in~\cite{Dhe} as parenthesized braids, an refinement of standard Artin braids in which one takes into account the distances between the strands. The above results show that, in the context of the $\ALD$-laws, the connection between~$\GALD$ and~$\Bp$ is similar to the connection between the geometry group of self-distributivity and the braid group~$B_\infty$, as investigated in~\cite{Dgd}. The benefit of the geometry monoid approach is to make the operations~\eqref{E:NNewop} natural and explain why they had to appear in this form and in this group.

\subsection{Presentation of $\GGLL$}

A different use of a blueprint is to allow for a closer comparison between the monoid~$\GGLL$ and the group~$\GLL$ in the non-linear case. We observed that $\GLL$ is constructed by means of a list of confluence relations holding in~$\GGLL$, but, in general, there is no reason why this list should be complete, \ie, provide a presentation of~$\GGLL$. Actually, when the empty operator belongs to~$\GGLL$, it cannot be the case that the confluence relations exhaust all relations of~$\GGLL$ as no such relation involves~$\emptyset$. However, the true question is to control the relations of~$\GGLL$ that do not involve the empty operator, which amounts to describing the relation~$\sim$ of Definition~\ref{D:Sim} on~$\GGLL$. Then, we have the following solution: 

\begin{prop}
\label{P:Pres}
Assume that $\phi$ is injective and $\Int$ is a $\phi$-blueprint for the laws~$\LLL$. Then the confluence relations used in the definition of~$\GLL$ generate all non-trivial relations in~$\GGLL$ in the following sense: for all word~$\ww, \ww'$ in~$\WLL$, the relation $\eev(\ww) \sim \eev(\ww')$ holds only if $\ww$ and $\ww'$ represent the same element of the group~$\GLL$.
\end{prop}

\begin{proof}
Assume that $\ww, \ww'$ are words in~$\WLL$ such that the associated operators~$\eev(\ww)$ and~$\eev(\ww')$ in~$\GGLL$ are connected by~$\sim$, \ie, there exists at least one term~$\tt$ on which they agree. Let~$\ttt$ be the common image of~$\tt$ under these operators. The hypothesis that $\Int$ is a $\phi$-blueprint gives
$$\Int(\tt) \cdot \phi(\ev(\ww)) = \Int(\ttt) = \Int(\tt) \cdot \phi(\ev(\ww'))$$
in~$\GLL$, whence $\phi(\ev(\ww)) = \phi(\ev(\ww'))$. If $\phi$ is injective, we deduce $\ev(\ww) = \ev(\ww')$, \ie, the words~$\ww$ and~$\ww'$ represent the same element of the group~$\GLL$.
\end{proof}

\begin{exam}
\label{X:Rel}
We conjecture that the endomorphism~$\sh0$ is injective on the group~$\GALD$, which would imply that the confluence relations listed in Section~\ref{SS:ConfALD} generate, in the sense described above, all relations holding in the geometry monoid~$\GGALD$. 

When we restrict from~$\ALD$ to~$\LD$ alone, then the argument is similar, and, in that case, the injectivity of~$\sh0$ on~$\GLD$ is known. What makes the case of~$\ALD$ more difficult is that, in the latter case, some confluence relations are missing, and the group~$\GALD$ is not a group of fractions for the positive monoid~$\GALDp$, while $\GLD$ is a group of fractions for~$\GLDp$. In both cases, proving the injectivity of~$\sh0$ on the positive monoid is easy, but extending the result to the group is not obvious hen the latter is not a group of fractions.
\end{exam}
 
\subsection{Solving the word problem}

Still another possibility is to use the blueprint for solving the word problem of~$\LLL$, \ie, to construct an algorithm that decides whether two terms~$\tt, \ttt$ are equivalent modulo the laws of~$\LLL$.

\begin{prop}
\label{P:WP}
Assume that $\Int$ is a Turing computable $\phi$-blueprint for the laws~$\LLL$, that $P, Q$ are disjoint recursively enumerable subsets of~$\GLL$ such that $P$ includes the image of~$\phi$, and there is a binary relation~$\rel$ on~$\TSx$ such that, for all terms~$\tt, \ttt$, at least one of~$\tt \EQ\LLL \ttt$, $\tt \rel \ttt$ holds and $\tt \rel \ttt$ implies $\Int(\tt)\inv \Int(\ttt) \in Q$. Then the word problem of the laws~$\LLL$ is solvable.
\end{prop}

\begin{proof}
Let $\widehat P, \widehat Q$ be disjoint recursive sets that include~$P$ and~$Q$ respectively. First assume that $\tt \EQ\LLL \ttt$ holds. Under the hypotheses, this implies 
$$\Int(\tt)\inv \Int(\ttt) \in \Im(\phi) \subseteq P \subseteq \widehat P.$$
Conversely, assume that $\tt \EQ\LLL \ttt$ fails. Then necessarily $\tt \rel \ttt$ holds, 
which implies
$$\Int(\tt)\inv \Int(\ttt) \in Q \subseteq \widehat Q,$$
so $\tt \EQ\LLL \ttt$ is equivalent to $\Int(\tt)\inv \Int(\ttt) \in \widehat P$. As the function~$\Int$ is assumed to be Turing computable and the set~$P$ is assumed to be Turing decidable, the latter condition is Turing decidable.
\end{proof}

\begin{exam}
Let us consider the case of the $\LD$-law. We know that there exists an $\sh0$-blueprint~$\Int$ for~$\LD$. Let $\tt \rel \ttt$ be the symmetric closure of the relation ``$\tt$ is $\LD$-equivalent to some iterated left subterm of some term $\LD$-equivalent to~$\ttt$''. Let $P$ be the subset of~$\GLD$ consisting of those elements that can be expressed using none of~$\SS\ea, \SS\ea\inv$, and let~$Q$ be the subset of~$\GLD$ consisting of those elements that can be expressed using exactly one of~$\SS\ea, \SS\ea\inv$. It is easy to show that every element of~$\GLD$ in the image of~$\sh0$ belongs to~$P$, while $\tt \rel \ttt$ implies $\Int(\tt)\inv \Int(\ttt) \in Q$. Moreover, one can show that $P$ and $Q$ are disjoint and recursive, so Proposition~\ref{P:WP} implies that the word problem of~$\LD$ is solvable \cite{Dgd}. This was the first solution for a long standing open question---other solutions are known now.

A similar scheme was used in~\cite{Dgj} to solve the word problem for the central duplication law $x(yz) = (xy)(yz)$---and, in this case, no alternative solution is known so far.

As for the word problem of the~$\ALD$-laws---for which a direct solution involving the group~$\Bp$ of Definition~\ref{D:Braids} is known---the scheme might work as well, but, as in Example~\ref{X:Rel}, some pieces are missing as the group~$\GALD$ fails to be a group of fractions for the positive monoid~$\GALDp$, making the verification of certain technical conditions problematic, typically the fact the expected sets~$P, Q$ are disjoint.
\end{exam}

\section{Proving global confluence}
\label{S:Conf}

Let $\LLL$ be a family of oriented algebraic laws. Then we have introduced both the geometry monoid~$\GGLL$, which corresponds to the rewrite system~$\RLL$ and to using the laws of~$\LLL$ with either orientation, and the positive geometry monoid~$\GGLLp$, which corresponds to the rewrite system~$\RLLp$ and to using the laws of~$\LLL$ in the distinguished direction only. An important case is when the system~$\RLLp$ turns out to be confluent, and it is a technically significant issue to prove this confluence result when possible.

The standard way for proving a confluence result consists in checking local confluence and then using some noetherianity condition to conclude using the classical Newman lemma---see for instance~\cite{Boo, Hue}. However, it turns out that, in the current framework, the rules one considers are often ill-oriented, so that no noetherianity can be expected. The aim of this short section is to present an alternative method that can be used instead.

\subsection{Least common expansion}

Let us start with the example of the $\LD$-law $x \op (y \op z) = (x\op y)\op (x\op z)$. If we orient it in the contracting direction $(x\op y)\op (x\op z) \to x\op (y\op z)$, then the rule diminishes the size and is therefore neotherian, but it is easily checked that confluence fails. When one chooses the expanding orientation, namely $x\op (y\op z) \to (x\op y)\op (x\op z)$, then, as seen in Section~\ref{SS:ConfALD}, we obtain a locally confluent system as, for each pair of addresses~$\a, \b$, there exists one (and exactly one) confluence relation $\SSp\a \comp ... = \SSp\b \comp ...$, where $\SSp\a$ denotes ``applying $\LD$ at position~$\a$ in the expanding direction''. Now, the rule increases the size of the terms, and it is not noetherian as, for instance, starting with the term $x \op (x \op x)$, we can apply~$\SSp\ea$ any number of times. So the general question is: 
\begin{quote}
How to prove that the locally confluent system~$\RLLp$ is possibly globally confluent (without assuming any noetherianity condition)?
\end{quote} 
Let us say that a term~$\ttt$ is a {\it degree~$d$} $\LLL$-expansion of a term~$\tt$ is $\ttt$ is the image of~$\tt$ under an element of~$\GGLLp$ that can be expressed as the product of at most~$d$ elementary operators~$\OOOp\LL\a$. The method developed in~\cite{Dgd} uses the following criterion:

\begin{prop}
Assume that there exists a mapping~$\partial: \TS \to \TS$ such that, for each term~$\tt$, the term~$\partial\tt$ is an $\LLL$-expansion of all degree~$1$ $\LLL$-expansions of~$\tt$ and, moreover, the mapping~$\partial$ is increasing w.r.t.~$\toLLp$, \ie, $\tt \toLLp \ttt$ implies $\partial\tt \toLLp \partial\ttt$. Then $\RLLp$ is confluent.
\end{prop}

\begin{proof}
It is enough to prove that, for each~$d$, the term $\partial^d\tt$ is a common $\LLL$-expansion of~$\tt$ of all degree~$d$ $\LLL$-expansions of~$\tt$. We use induction on~$d$. The result is trivial for $d = 0$, and it is true for $d = 1$ by hypothesis. Assume $d \ge 2$, and let $\ttt$ be a degree~$d$ $\LLL$-expansion of~$\tt$. By hypothesis, there exists~$\tti$ such that $\tti$ is a degree~$d-1$ expansion of~$\tt$, and $\ttt$ is a degree~$1$ expansion of~$\tti$. On the one hand, $\ttt$ is a degree~$1$ expansion of~$\tti$, hence $\partial \tti$ is an expansion of~$\ttt$. On the other hand, by induction hypothesis, $\partial^{d-1}\tt$ is an expansion of~$\tti$, hence $\partial^d\tt$ is an expansion of~$\partial\tti$. Being an expansion is transitive, hence $\partial^d\tt$ is an expansion of~$\ttt$, as expected. 
\end{proof}

The criterion was first designed for the case of self-distributivity \cite{Dgd}, but it was subsequently also applied in the case of central duplication \cite{Dgj}, and in the case of idempotency with or without self-distributivity \cite{Lrb}. In each case, the construction of the operator~$\partial$ heavily relies on the considered laws.

\subsection{Group of fractions}

An alternative solution for proving the possible confluence of the positive rewrite system~$\RLLp$ consists in working in the geometry monoid~$\GGLL$, or, rather, in its positive submonoid~$\GGLLp$. Then we have the following criterion:

\begin{prop}
\label{P:Mult}
Assume that the monoid~$\GGLLp$ admits common right multiples in the following strong sense: for all~$\ff, \gg$ in~$\GGLLp$, there exist $\ff', \gg'$ satisfying $\ff \comp \gg' = \gg \comp \ff'$ and, in addition, the domain of~$\ff \comp \gg'$ is the intersection of the domains of~$\ff$ and~$\gg$. Then the rewrite system~$\RLLp$ is confluent.
\end{prop}

\begin{proof}
Write $\tt \toLLp \ttt$ if $\tt$ rewrites to~$\ttt$ with respect to~$\RLLp$. Now assume $\tt \toLLp \ttt$ and $\tt \toLLp \tttt$. By Proposition~\ref{P:Basic}, there exist~$\ff, \gg$ in~$\GGLLp$ satisfying $\ttt = \tt \act \ff$ and $\tttt = \tt \act \gg$. By Proposition~\ref{P:Mult}, there exist~$\ff', \gg'$ in~$\GGLLp$ satisfying $\ff \comp \gg' = \gg \comp \ff'$ and such that the domain of $\ff \comp \gg'$ is the intersection of the domains of~$\ff$ and~$\gg$, hence contains~$\tt$. Then we have
$$\ttt \act \gg' = \tt \act (\ff \comp \gg') = \tt \act (\gg \comp \ff') = \tttt \act \ff',$$
\ie, letting $\ttttt = \tt \act (\ff \comp \gg')$, we have $\ttt \toLDp \ttttt$ and $\tttt \toLDp \ttttt$.
\end{proof}

In good cases, Proposition~\ref{P:Mult} can be proved using the group~$\GLL$, or rather the positive monoid~$\GLLp$, and its presentation: here we denote by~$\GLLp$ the monoid generated by  generators~$\OOO\LL\a$ for~$\LL$ in~$\LLL$ and~$\a$ an address, subject to all confluence relations connecting the operators~$\OOOp\LL\a$ in~$\GGLL$, \ie, the monoid that admits, as a monoid, the same presentation as the group~$\GLL$. Optimally, we could expect that the possible existence of common right multiples in~$\GLLp$ implies the existence of common right multiples in~$\GGLLp$, in the strong form required in Proposition~\ref{P:Mult}. Always because of the empty operator, this need not be the case in general. Nevertheless, we have

\begin{prop}
Assume that $\LLL$ consists of semi-linear laws, \ie, laws $\tm \to \tp$ with $\tm$ an injective term (no variable repeated), and $\GLLp$ admits common right multiples. Then $\GGLLp$ admits common right multiples and Proposition~\ref{P:Mult} applies.
\end{prop}

\begin{proof}[Sketch of proof]
Under the hypotheses, the empty operator does not appear in~$\GGLLp$, and it follows that $\GGLLp$ is a homomorphic image of~$\GLLp$. Thus every relation in~$\GLLp$ projects into a relation in~$\GGLLp$, and, in particular, common multiples in~$\GLLp$ induce common multiples in~$\GGLLp$.
\end{proof}

So the question remains of proving the existence of common right multiples in the monoid~$\GLLp$. By hypothesis, we start with a presented monoid where the relations
are of the form $\OOO\LL\a \cdot ... = \OOO\LLbis\b \cdot ...$, \ie, they assert the existence of some common right multiples, and what we need is to extend the result so as to show that any two elements of the monoid admit a common right multiple. This is easy if all relations in the presentation turns out to have length at most~$2$ for, in that case, an easy induction gives a common multiple of length at most~$p+q$ for any two elements that can be expressed by words of length~$p$ and~$q$. In more complicated cases---typically in the case of the $\LD$-law, where we have seen some confluence relations involve words of length~$3$ or~$4$, like $\SS\ea \SS1 \SS\ea = \SS1 \SS\ea \SS1 \SS0$---the method consists in finding a family of words~$X$ that includes all generators~$\OOO\LL\a$ and is closed under complement in the sense that, for all words~$u, v$ in~$X$, there exists $u', v'$ in~$X$ such that both $uv'$ and~$vu'$ represent a common right multiple of the elements of~$\GLLp$ represented by~$u$ and~$v$. For instance, this approach works for~$\LD$, but it is rather delicate and leads to an upper bound which, instead of~$p+q$ as above, is a tower of exponentials of height~$2^{p+q}$~\cite{Dgd}.

When one can prove that common (right) multiples exist in the monoid~$\GLLp$, and, that, in addition, the latter admits cancellation, then standard results \cite{CPr} guarantee that the group~$\GLL$ is a group of fractions for~$\GLLp$, and this turns out to be crucial in the study of~$\GLL$. Again, proving that $\GLLp$ is cancellative requires specific tools connected with the so-called word reversing method~\cite{Dgp}. In many good cases (self-distributivity, associativity, central duplication), one can show that the divisibility relation gives~$\GLLp$ the structure of a lattice---but this is beyond the scope of this paper.

\section{Summary}
\label{S:Summary}

We thus showed how introducing by means of confluence relations an abstract group that is supposed to mimick the properties of the geometry monoid---hence of the initial rewrite system---and then internalizing terms in that group so as to transform the initial external action into an internal multiplication may allow to solve nontrivial questions about a given family of algebraic laws. 

In the case of non linear algebraic laws, \ie, when some variable is repeated at least twice, due to the problem of the empty operator, the geometry monoid is an intrinsically inconvenient object, and our approach for replacing it with a group is the only known one. Of course, one might consider other relations than the confluence relations. The latter proved to be suitable for the examples considered here, but different schemes might prove relevant for other identities: for instance, when commutativity is involved, the operators are involutive, and the associated relations $\OOO\LL\a^2 = 1$ are not confluence relations. In some cases, one can keep the principle of introducing the structure~$\GLL$ presented by the confluence relations in~$\GGLL$, but taking~$\GLL$ to be a monoid rather than a group: typically, this has to be done in the case of the {\it idempotency law} $x = xx$, as, in this case, confluence relations of the type
$\OOOp\LL\ea \comp \OOOp\LL\ea = \OOOp\LL\ea \comp \OOOp\LL0 \comp \OOOp\LL1$ are satisfied, preventing $\GGLL$ from admitting left cancellation---see~\cite{Jed}. So we see that the methods described here require some flexibility in their application.

However, it should be possible to adapt all three main steps, namely introducing a monoid of partial operators, replacing it with a group using a presentation, and internalizing the rewrite system by representing the objects on which the initial action is defined by copies inside the group, to more general frameworks. An example is given in~\cite{Dhb} where algebraic laws are replaced with a more complicated action on terms (``twisted commutativity''); we think that more rewrite systems, possibly of a completely different type, could be investigated using similar tools.

\end{document}